\documentclass[aps,prl,floatfix,twocolumn,showpacs,amsmath,amssymb,superscriptaddress,titlepage,natbib]{revtex4-1}
\usepackage{graphicx}% Include figure files
\usepackage{epstopdf}
\usepackage{epsfig}
\usepackage{dcolumn}% Align table columns on decimal point
\usepackage{bm}% bold math
\usepackage{color}
\usepackage{subfigure}
\usepackage{wrapfig}
\usepackage{rotate,color,url}
\usepackage{longtable}
\usepackage{time}
\usepackage{adjustbox}

\usepackage[pagebackref=false,colorlinks,linkcolor=blue,citecolor=blue,urlcolor=blue]{hyperref}
\usepackage{multirow,varwidth}
\newcommand{\bmno}{Bi$_3$Mn$_4$O$_{12}$(NO$_3$) }

\begin{document}
\title{Origin of magnetic frustration in \bmno}

\author{Mojtaba Alaei}
%\email{m.alaei@cc.iut.ac.ir}
\affiliation{Department of Physics, Isfahan University of
Technology, Isfahan 84156-83111, Iran.}

\author{Hamid Mosadeq}
\affiliation{Department of Physics, Faculty of Science, Shahrekord University, Shahrekord 88186-34141, Iran.}
\author{Ismaeil Abdolhosseini Sarsari}
\affiliation{Department of Physics, Isfahan University of
Technology, Isfahan 84156-83111, Iran.}
\author{Farhad Shahbazi}
\email{shahbazi@cc.iut.ac.ir}
\affiliation{Department of Physics, Isfahan University of Technology, Isfahan 84156-83111, Iran.}
%\affiliation{Department of Physics and Astronomy, University of Waterloo, Ontario, N2L 3G1, Canada}

%\author{S. Javad Hashemifar}
%\affiliation{Department of Physics, Isfahan University of
%Technology, Isfahan 84156-83111, Iran.}
\date{\today}

\begin{abstract}
\bmno (BMNO) is a honeycomb bilayers anti-ferromagnet, not showing any ordering down to very low temperatures despite having a relatively large Curie-Weiss temperature. 
Using ab initio density functional theory, 
we extract an effective spin Hamiltonian for this compound. 
The proposed spin Hamiltonian consists of anti-ferrimagnetic Heisenberg terms with coupling constants 
ranging up to third intra-layer and fourth inter-layer neighbors. Performing Monte Carlo simulation, 
we obtain the temperature dependence of magnetic susceptibility and so the Curie-Weiss temperature and 
find the coupling constants which best matches with the experimental value. 
We discover that depending on the strength of the interlayer exchange couplings,  two collinear spin configurations 
compete with each other in this system. Both states have in plane N{\'e}el character, 
however, at small interlayer coupling spin directions in the two layers are antiparallel (N$_1$ state) 
and discontinuously transform to parallel (N$_2$ state) by enlarging the interlayer couplings 
at a first order transition point.   Classical Monte Carlo simulation and density matrix renormalization 
group calculations confirm that exchange couplings obtained for BMNO are in such a way that put 
this material {at  the  phase boundary of a first order phase transition}, where the trading between 
these two collinear spin states prevents it from setting in a magnetically ordered state. 
\end{abstract}


\maketitle

\begin{figure}[t]
   \includegraphics[width=0.9\columnwidth]{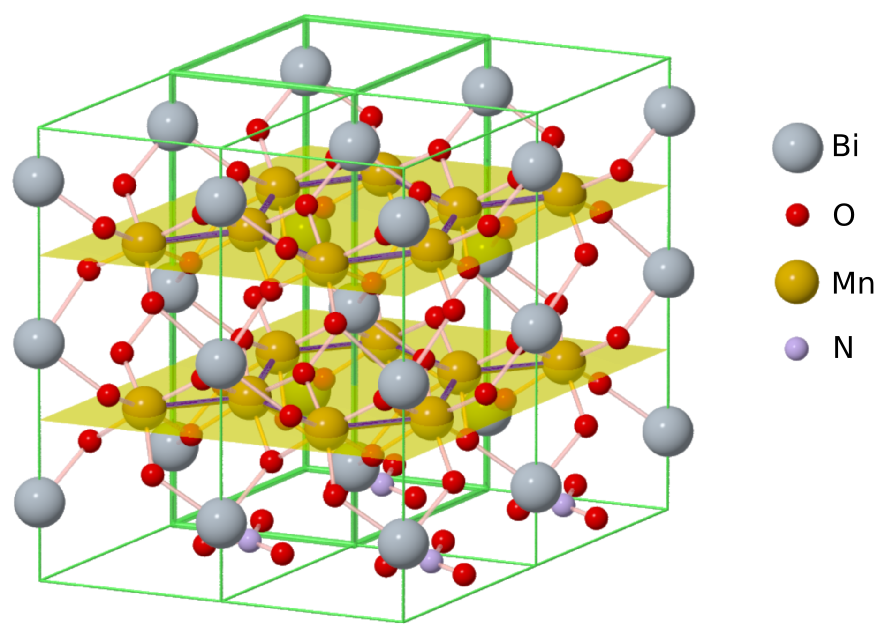}
    \caption{(Color online) The $2\times2$ supercell of BMNO. The thicker green lines show the 
             primitive cell. Two (yellow transparent) planes show honeycomb lattices 
             which are made of Mn atoms. }
    \label{fig:bmno}
\end{figure}

\bmno (BMNO) is an experimental realization of frustrated honeycomb magnetic materials, synthesized by Smirnova {et al.}~\cite{bmnostruct}. In this compound, the magnetic lattice can be effectively described by a weakly coupled honeycomb bilayers of Mn$^{+4}$ ions (Fig.~\ref{fig:bmno}). 
The temperature dependence of magnetic susceptibility of BMNO does not indicate any ordering down to $T=0.4$K, in spite of the Curie-Weiss temperature  $\theta_{\rm CW}\approx -257 $K~\cite{bmnostruct,masudaprb}. The absence of long-range ordering in BMNO is also confirmed by specific heat measurements~\cite{bmnostruct,masudaprb}, neutron scattering~\cite{matsudaprl} and high-field electron spin relaxation (ESR) experiments~\cite{ESR}. 
So far, the theoretical attempts to explain the magnetic properties of BMNO have been focusing on the frustration effect of second intra-layer coupling $J_2$ or the  tendency toward dimerization by considering a large 
anti-ferromagnetic inter-layer nearest neighbor  coupling $J_{1c}$
~\cite{kandpalprb,ganesh2011quantum,ganesh2011neel,oitmaa2012ground,dmbmno,
zhang2014quantum,albarracin2016field,bishop2016frustrated}.  
{ In an attempt to calculate the exchange interactions by ab initio method, it is found that the dominant exchange interactions are  nearest in-plane coupling  ($J_1$) and also an effective inter-plane coupling ($J_{c}$)  which exceeds $J_1$~\cite{kandpalprb}. However, in that work the experimental positions of the atoms in the structure are taken without geometry optimization and only $5$ magnetic configurations are considered for the  calculation of exchange interactions.  }
In this paper, we obtain a Heisenberg spin Hamiltonian for BMNO, using an ab initio LDA+U  calculation. 
 In our calculations, we consider a detailed analysis of nonidentical Mn atoms which were assumed to be identical in the previous calculation~\cite{kandpalprb}. 
We show that how this consideration can affect the 
exchange couplings in the spin Hamiltonian. We find that in contrary to the previous works, none of $J_2$ and $J_{1c}$ are large enough to  frustrate BMNO to reach an ordered state. Indeed, surprisingly  the interlayer coupling constants are fined tuned in a way that make this system living at the edge boundary of two competing magnetic states. 
%Instead, it is the second neighbor inter-layer coupling $J_{2c}$ which has a crucial role in determining the magnetic properties of this compound.  

%%%%%%%%%%%%%%%%%%%%%%%%%%%FIG1
%%%%%%%%%%%%%%%%%%%%%%%%%%%FIG1

%%%%%%%%%%%%%%%%%%%%%%%%%%%%%%FIG2
\begin{figure}
   \includegraphics[width=0.7\columnwidth]{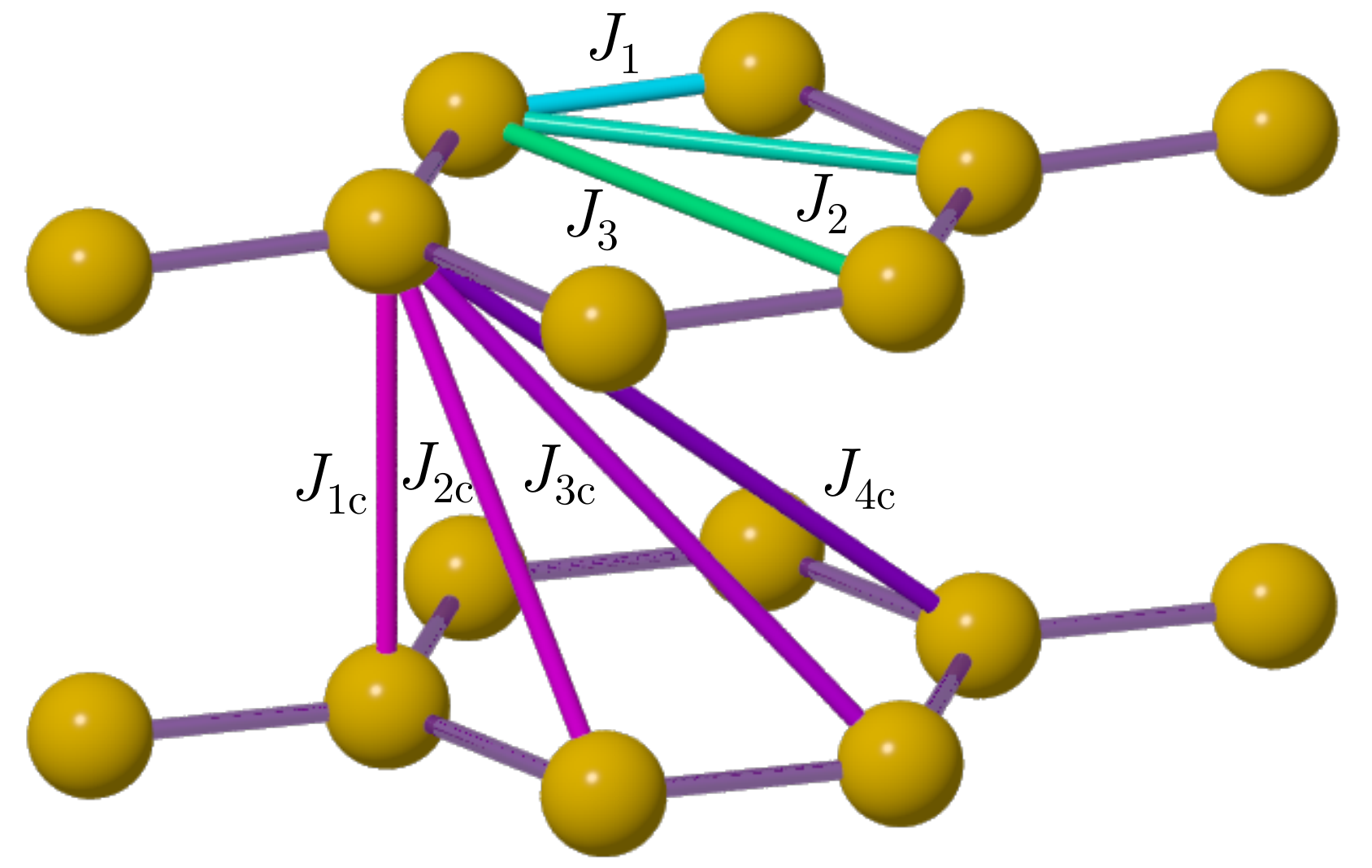}
    \caption{$J_1$, $J_2$ and $J_3$ indicate the Heisenberg exchange coupling constants between first, second and third nearest neighbor in-plane Mn$^{+4}$ ions, respectively. $J_{1c}$, $J_{2c}$, $J_{3c}$ and $J_{4c}$ indicate the Heisenberg exchange coupling constants between inter-plane first, second and third nearest neighbor Mn$^{+4}$ ions, respectively. }
    \label{fig:Jij}
\end{figure}
%%%%%%%%%%%%%%%%%%%%%%%%%%%%%FIG2
{\em Ab initio method}.
To derive magnetic exchange couplings, we employ Density Functional Theory (DFT) 
with Full-Potential Local-Orbital minimum-basis (FPLO), using  FPLO code~\cite{FPLO} (FPLO14.00-45). 
For charge analysis we employ Projector Augmented Wave (PAW) method, using  Quantum-Espresso (QE) distribution~\cite{QE}. 
To account for exchange-correlation interaction we use PBE functional~\cite{PBE} 
from Generalized Gradient Approximation (GGA). To improve estimation 
of electron-electron Coulomb interactions,  we also add Hubbard-like $U$ correction to DFT calculations, i.e.,  
DFT+$U$~\cite{anisimov1991,anisimov1993}. 
To implement DFT+$U$, FPLO uses Liechtenstein's approach~\cite{LDAU,FPLOLDAU}. 
In Liechtenstein's approach the two parameters, $U$ (on-site Coulomb repulsion)
and $J_H$ (the on-site Hund exchange) needs to be set, which we use $J_H=1.0$ eV and $U=1.5, 2.0, 3.0$ and $4.0$ eV.

%!!!!!!!!!!!!!!!!!!!!!!!!!!!!FIG3
\begin{figure}[t]
   \includegraphics[width=\columnwidth]{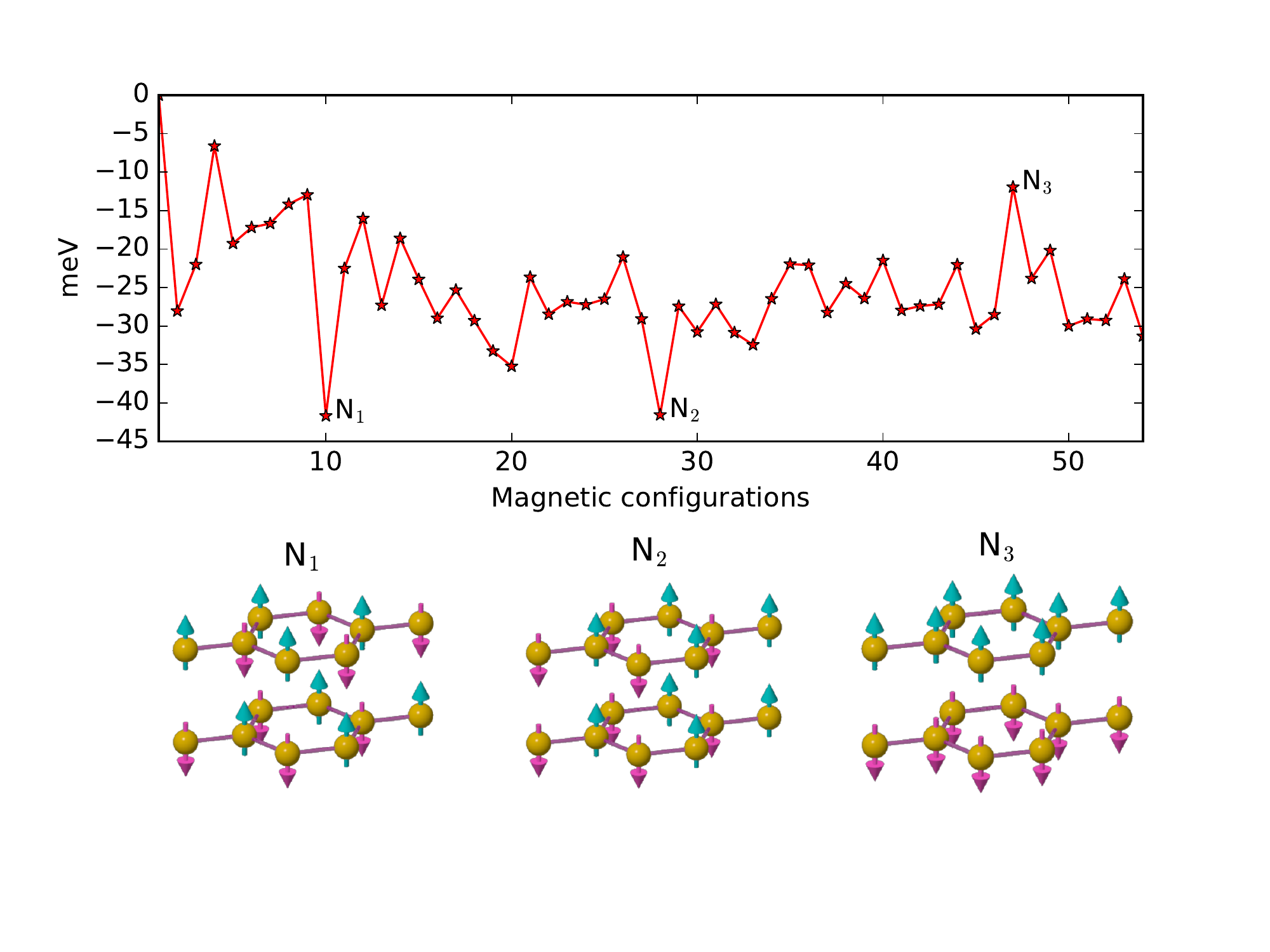}
    \caption{ (Color online) ({\bf {top}}): The ab initio energy landscape (per Mn atom) of $54$ magnetic configurations obtained by DFT+$U$ with $U=1.5$  
                   eV. The energies of magnetic configurations are respect to fully ferromagnetic state whose energy is set to $0$.  
                   ({\bf {bottom}}):  Three magnetic configurations N$_{1}$, N$_{2}$  and N$_{3}$. }
 \label{fig:configs}
\end{figure}
%!!!!!!!!!!!!!!!!!!!!!!!!!!!!FIG3

{\em Spin Hamiltonian}.
The strategy of finding an effective spin Hamiltonian from ab initio calculations is to first compute the ground state energy for some given magnetic configurations. Then, mapping the energy difference of these configurations to an appropriate spin model gives us the coupling constants of the model. In this work, we use non-relativistic DFT, hence any magnetic anisotropy originating from the spin-orbit interaction is ignored in this approximation. Therefore, to leading order, we propose a spin Hamiltonian containing only bi-linear Heisenberg interactions,  
%\begin{equation}
%\label{H}
${\cal H}_{\rm{Heisenberg}}=\sum_{i>j} J_{ij} \, {\bf n}_i \cdotp {\bf  n}_j$, 
%\end{equation}
where ${\bf  n}_i$ and ${\bf  n}_j$ are classical unit vectors representing 
the orientation of the magnetic moments at sites $i$ and $j$, respectively, 
with exchange interactions $J_{ij}$ between them. 
The primitive cell of \bmno 
contains $23$ atoms.  Because of the limitation in  computational resources,
we use the $2 \times 2$ supercell containing $92$ atoms (Fig.~\ref{fig:bmno}). This lets us  
calculate $J_{ij}$'s up to the third   in-plane  neighbor
($J_1$, $J_2$ and $J_3$) and up to the fourth   inter-plane neighbor coupling 
($J_{1c}$, $J_{2c}$, $J_{3c}$ and $J_{4c}$) (see figure \ref{fig:Jij}). 
BMNO is metallic in  GGA, however, implementing   
spin-polarized calculation makes this compound  insulating,  independent of its magnetic configuration. 
Within co-linear spin polarized  GGA, the ground state is a N{\'e}el state in which the nearest neighbor Mn magnetic 
moments (in and out of plane) are anti-parallel with respect to each other. This magnetic configuration is marked by N$_{1}$ in Fig.~\ref{fig:configs}. 
We calculated the total energy  for more than $50$ independent magnetic configurations. Then employing the least square method, enables us to obtain the exchange couplings with the accuracy of $0.02$ meV. The top panel of figure \ref{fig:configs} represents the energy landscape (per Mn atom) calculated for $54$ different magnetic configurations within the super-cell shown in Fig.~\ref{fig:bmno}. The detailed description of these configurations is given in Ref.\cite{supplement}. As it is obvious from this figure, the two configurations N$_{1}$ and N$_{2}$ (bottom panel of Fig.~\ref{fig:configs}) are very  close in energy space. In configuration N$_{2}$, the magnetic ordering in each honeycomb layer is  N{\'e}el type, but unlike N$_{1}$, the magnetic moment orientations of two layer are parallel. 
The coupling constants of the Heisenberg Hamiltonian, obtained by different values of on-site Coulomb interaction $U$,  are given in  Table~\ref{tab:J-relax}. {We also checked that the exchange interactions between the adjacent  honeycomb bilayers are negligible comparing to the ones  inside the bilayers~\cite{supplement}.}   
It is important to mention that to achieve equal couplings between equivalent Mn ions in the two layers, we need to geometrically optimize the atomic positions rather than just using the experimental atomic positions (for the details see supplementary information~\cite{supplement}).   

%table2!!!!!!!!!!!!!!!!!!!!!!!!!!!!!!!!!!!!!!!!!!!!!!!!!!!!!!!!!!!!!!!!!!!!!!!!!!!!!!!!!!!!!!!!!!!!!!!!!!!
\begin{table*}
\caption{Heisenberg constants  obtained  by ab initio calculations (LDA+$U$) using 
different $U$. The geometrically optimized structure 
(in the ferromagnetic state)  is used in these calculations. The last column shows 
Curie-Weiss temperature obtained from Monte Carlo simulations 
for a system size with $N=4\times 24 \times 24 \times 1$. 
The experimental Curie-Weiss temperature  is between $-257$K~\cite{bmnostruct} and
$-222$K~\cite{masudaprb}.
}
\label{tab:J-relax}
\begin{ruledtabular}
\begin{tabular}{c|c|cccccccc}
method & $U ({\rm eV})$   &  $J_1\, ({\rm meV})$ & $J_2\, ({\rm meV})$ & $J_3\, ({\rm meV})$ & $J_{1c}\, ({\rm meV})$  & $J_{2c}\, ({\rm meV})$ & $J_{3c}\, ({\rm meV})$ & $J_{4c}\, ({\rm meV})$ & $\Theta_{CW} (K)$\\ \hline 

\multirow{3}{*}{FPLO}                                    
& 1.5                       &  10.7               &      0.9            &        1.2       &   3.0                  & 1.1                     &      0.5               &  0.9   &  -244\\
 &  2.0                     &  9.0                 &      0.8            &        1.0       &   2.6                   &  0.9                   &      0.5               &  0.8  & -203  \\ %\hline
                                                          &  3.0                     &  6.6                 &      0.6            &        0.8       &   2.1                   &  0.7                   &      0.3               &  0.6  & -144 \\ %\hline
                                                          &  4.0                     &  5.1                 &      0.5            &        0.6       &   1.7                   &  0.6                   &      0.3               &  0.5  & -111  %\\ \hline 
\end{tabular}
\end{ruledtabular}
\end{table*}
%table2!!!!!!!!!!!!!!!!!!!!!!!!!!!!!!!!!!!!!!!!!!!!!!!!!!!!!!!!!!!!!!!!!!!!!!!!

%!!!!!!!!!!!!!!!!!!!!!!!!!!!!!!!!

%\section{Results}
{\em The Mn spin state}.
The bond valence sum  indicates the  valence state 
Bi$^{3+}_3$Mn$^{4+}_4$O$^{2-}_{12}$(NO3)$^-$ for BMNO~\cite{bmnostruct}. However,  
using the charge analyzing code  Critic2~\cite{critic2, critic}, within GGA/PAW, we find the valence state 
 Bi$^{1.96+}_3$Mn$^{1.87+}_4$O$^{1.04-}_{12}$(NO3)$^{0.86-}$ in N$_{1}$-configuration. 
This charge distribution will not change dramatically in the case of implementing  DFT+$U$ even 
with large U parameter. {We also made sure that using an all-electron method, such as FPLO,  this picture of charge distribution remains  nearly unchanged.} 
The  local density analysis (Lowdin charges), also proposes the charge distribution Bi$^{1.49+}_3$Mn$^{1.45+}_4$O$^{0.74-}_{12}$(NO3)$^{0.40-}$.
These charge analyses show that the Mn-O bonds are ionic-covalent
instead of being completely ionic. Indeed, the reason for such a fractional charge distribution in BMNO is the strong hybridization between Mn d-orbitals and the neighboring  O p-orbitals. This also lowers the magnetic moment of Mn ions from $3\mu_{B}$ to about $2.5\mu_{B}$~(see Table I in supplementary information~\cite{supplement}).

%%%%%%%%%%%%%%%%%%%%%%%%%%%FIG4
\begin{figure}[t]
   \includegraphics[width=1.0\columnwidth]{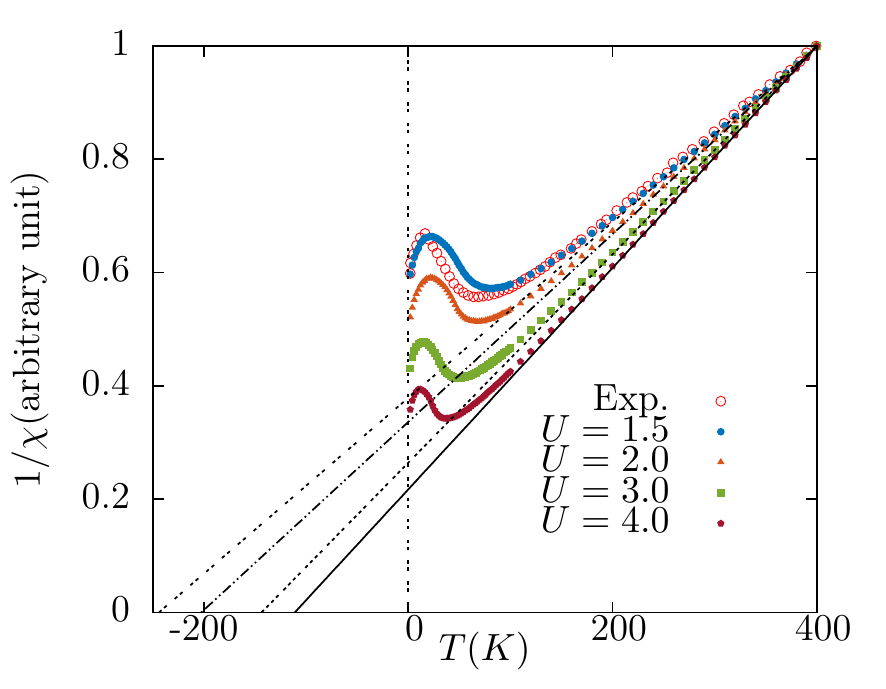}
    \caption{(Color online) Temperature dependence of  the inverse  normalized DC susceptibility ($1/\chi$) of BMNO 
    obtained in experiment and MC simulations for different set of ab initio exchange couplings 
    derived  by using  $U=1.5$ eV (filled circles), $U=2.0$ eV (triangles), $U=3.0$ eV (squares) and $U=4.0$ eV (pentagons).  The experimental data (empty circles) is extracted  from figure $7$ of Ref.~\cite{bmnostruct}. For the normalization,  all the data are divided  by their values  at $T=400$. 
    The crossing of the line fitted at high temperatures to $1/\chi$ with the horizontal axis gives the Curie-Weiss temperature.  }
    \label{fig:chi}
\end{figure}
%%%%%%%%%%%%%%%%%%%%%%%%%%%FIG4
%%%%%%%%%%%%%%%%%%%%%%%%%%%FIG5
\begin{figure}[b]
   \includegraphics[width=1.0\columnwidth]{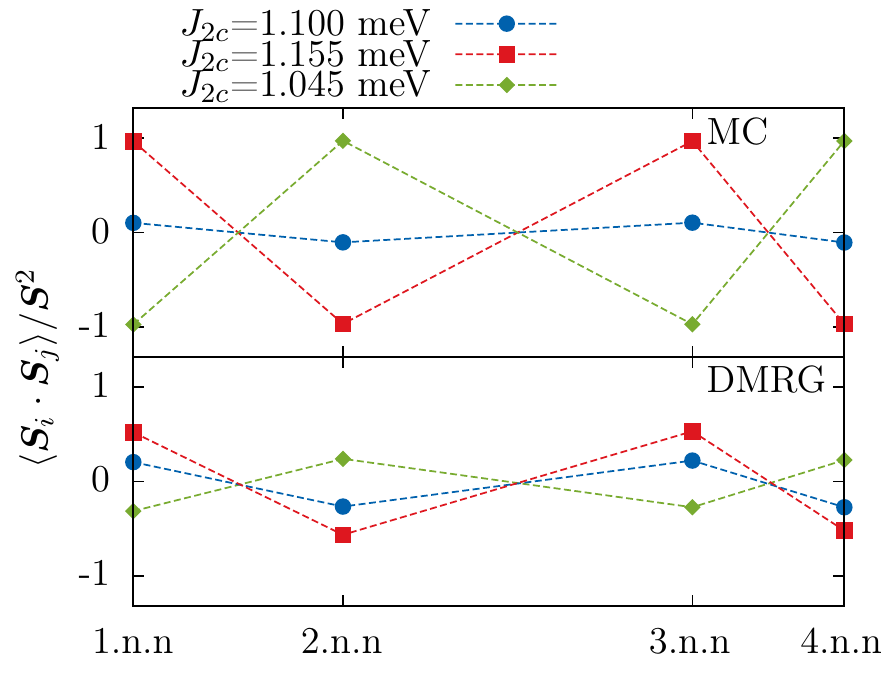}
    \caption{(Color online) Interlayer spin-spin correlation up to  neighbor obtained by ({\bf top}) MC simulation, ({\bf bottom} DMRG (normalized by $S^2$). $J_1,J_2,J_3,J_{1c},J_{3c}$ and $J_{4c}$ are kept fixed at those found by  $U=1.5$ eV. The spin-spin correlations at $J_{2c}=1.100$ meV are obtained by using $U=1.5$ eV (blue circles) and compared with $J_{2c}=1.045$ (green diamonds) and $J_{2c}=1.155$ meV (red squares).  }
    \label{fig:correlation}
\end{figure}
%%%%%%%%%%%%%%%%%%%%%%%%%%%FIG5

%%%%%%%%%%%%%%%%%%%%%%%%%%%%%%%%%%%%%%%%%%%%%%%%%%fig.6
\begin{figure}
\includegraphics[trim = 0mm 0mm 0mm 0mm, clip,width=1.0\columnwidth]{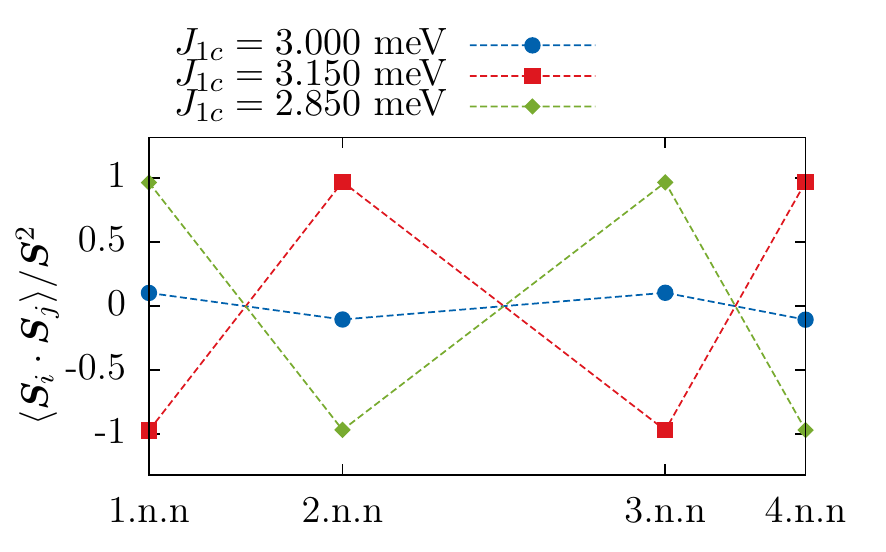}
\caption{(Color online) Interlayer spin-spin correlation (normalized by $S^2$) up to fourth neighbor obtain by MC simulation . $J_1,J_2,J_3,J_{2c},J_{3c}$ and $J_{4c}$ are kept fixed at the values obtained by  $U=1.5$ eV. The spin-spin correlations at $J_{2c}=3.0$ meV are obtained by using $U=1.5$ eV (blue circles) and compared with $J_{1c}=3.15$ (red squares) and $J_{1c}=2.85$ meV (green diamonds). }
\label{fig:corr2}
\end{figure}
%%%%%%%%%%%%%%%%%%%%%%%%%%%%%%%%%%%%%%%%%%%%%%%%%%
%%%%%%%%%%%%%%%%%%%%%%%%%%%%%%%%%%%%%%%%%%%%%%%%%%fig.7
\begin{figure}[t]
\includegraphics[trim = 0mm 0mm 0mm 0mm, clip,width=1.0\columnwidth]{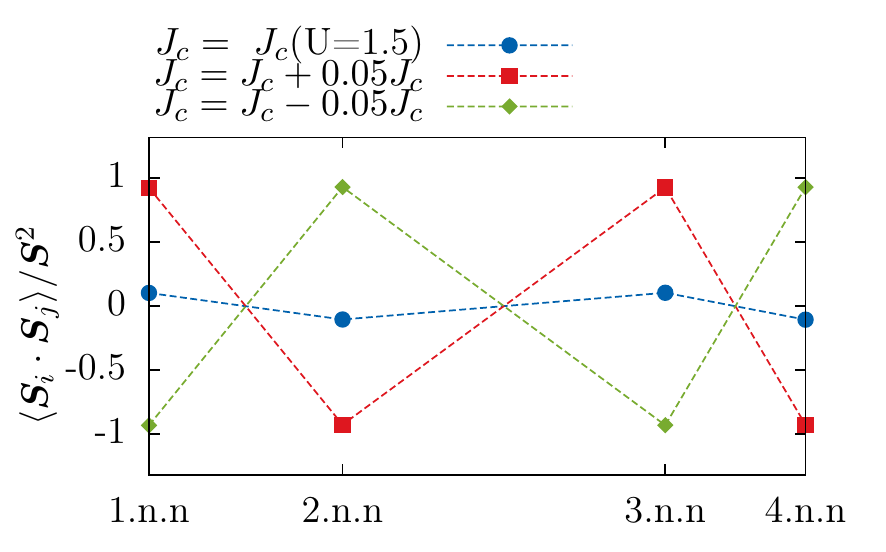}
\caption{(Color online) Interlayer spin-spin correlation (normalized by $S^2$) up to fourth neighbor obtain by MC simulation . $J_1,J_2,J_3$ are kept fixed at the values obtained using $U=1.5$ eV. The spin-spin correlations for the interlayer coupling calculated at $U=1.5$ (blue circles) are compared with the ones increased (red squares) and decreased (green diamonds) by 5 percent. }
\label{fig:corr3}
\end{figure}
%%%%%%%%%%%%%%%%%%%%%%%%%%%%%%%%%%%%%%%%%%%%%%%%%%

%%%%%%%%%%%%%%%%%%%%%%%%%%%%%%%%%%%%%%%%%%%%%%%%%%fig.8
\begin{figure}
\includegraphics[width=0.9\columnwidth]{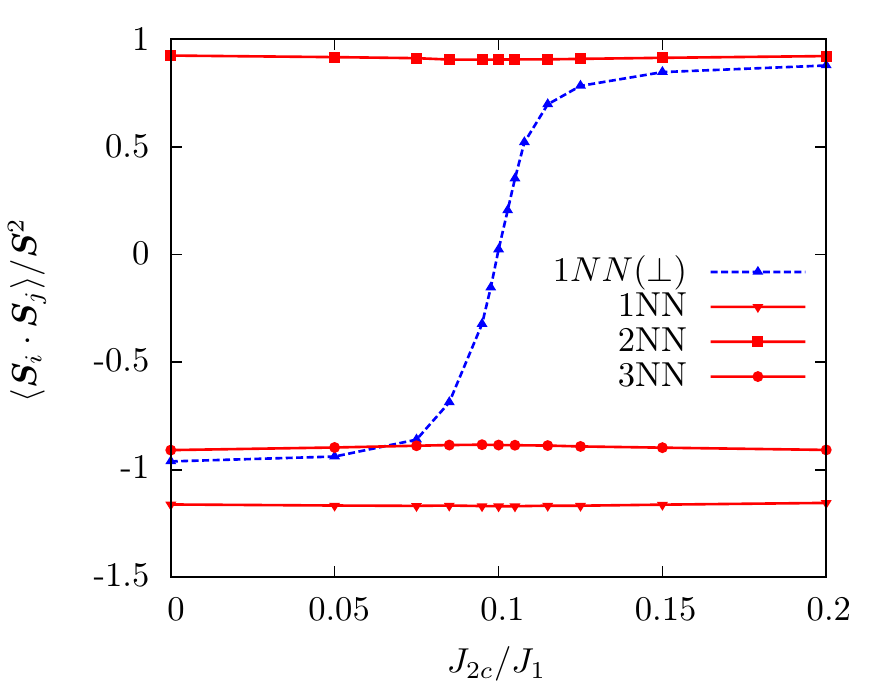}
\caption{(Color online) DMRG results for the variation of spin-spin correlations for first, second, third in-plane neighbors and first inter-plane neighbors versus   $J_{2c}/J_1$. Other coupling are fixed by those obtained by $U=1.5$ eV.}
\label{fig:dmrg}
\end{figure}
%%%%%%%%%%%%%%%%%%%%%%%%%%%%%%%%%%%%%%%%%%%%%%%%%%
%%%%%%%%%%%%%%%%%%%%%%%%%%%%%%%%%%%%%%%%%%%%%%%%%%fig.9
\begin{figure}
\includegraphics[trim = 5mm 0mm 0mm 0mm, clip,width=0.9\columnwidth]{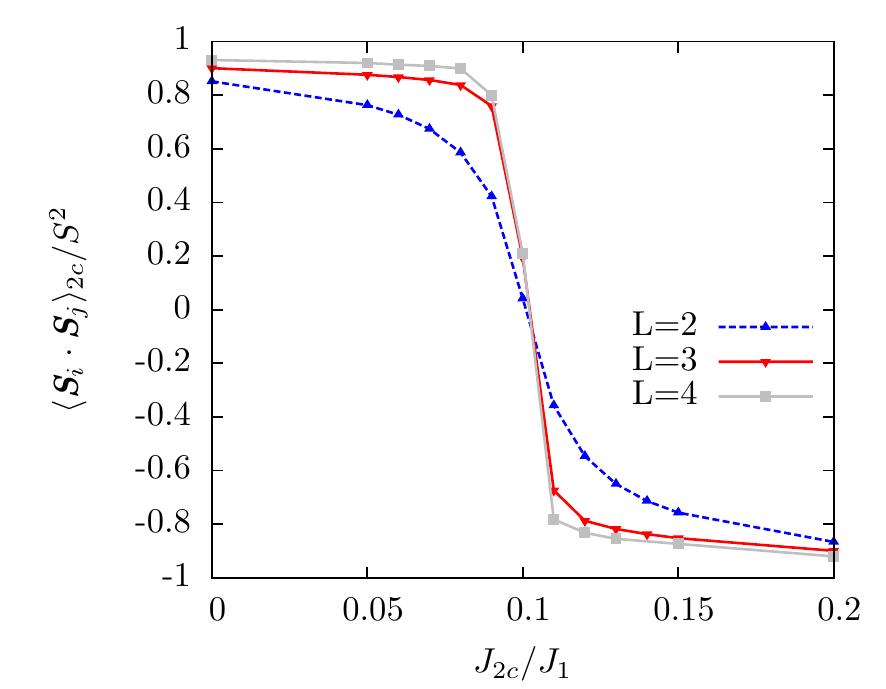}
\caption{(Color online) DMRG results for the variation of  the second interlayer spin-spin correlations  versus $J_{2c}/J_1$, for lattices of size $4\times L^2$ with  $L=2,3,4$. Other coupling are fixed by those obtained by $U=1.5$ eV.}
\label{fig:FH}
\end{figure}
%%%%%%%%%%%%%%%%%%%%%%%%%%%%%%%%%%%%%%%%%%%%%%%%%%

%!!!!!!!!!!!!!!!!!!!!!!!!!!!!!
{\em Monte Carlo Simulations}. 
To gain insight into the finite temperature properties of the model Hamiltonian,
we perform MC simulation. 
MC simulations are done  on a $24 \times 24 \times 1$ supercell
contains $2304$ Mn atoms with periodic boundary conditions.
We use single spin Metropolis updating, $3\times 10^6$ MC steps for  thermalization and 
$7 \times 10^6$ MC samplings for the measurement of physical quantity. To reduce the correlations,
we skip $5$ MC sweeps between successive data collections.    
{ In Figure \ref{fig:chi}  the temperature dependence of inverse  magnetic susceptibility for the sets of exchange 
couplings obtained by ab initio method using $U=1.5, 2.0, 3.0,4.0$  eV and also the experimental values are compared.
This figure shows a very good  agreement between the experimental values and the set of exchanges obtained by using $U=1.5$ eV}. 
 The linear fit at high temperatures crosses the $T$-axis
at a negative value which is the Curie-Weiss temperature $\theta_{CW}$. It can be seen that $\theta_{CW}$ increases by increasing the value of 
onsite Coulomb repulsion $U$. The $\theta_{CW}=-244 K$ is closest to what has been measured experimentally.     
To speculate about the ground state of the Hamiltonian, we calculated the inter-layer spin-spin correlation at a low temperature, up to fourth neighbor (Fig.~\ref{fig:correlation}). The correlations are calculated by averaging over $5\times 10^4$ MC samplings at $T=2K$. As it can be seen from this figure, for the couplings corresponding to $U=1.5$ eV  (Table I), the spin-spin correlations between the two layers are very small. We observe that, increasing (decreasing) the value of interlayer coupling by a little amount pushes the system toward N$_{1}$ (N$_{2}$) type ordering. Figure~\ref{fig:correlation} shows the change of  spin-spin correlation patterns, when  $J_{2c}$ varies by  only $\pm5$ percent, while keeping the rest of the couplings unchanged.  The same results are obtained when only $J_{1c}$  is changed while the other couplings are fixed (Fig.\ref{fig:corr2}) and also in the case that all the interlayer couplings are shifted up or downward  (Fig.\ref{fig:corr3}).  We also found  that the small changes in the in-plane couplings do not induce spin ordering in the system.     

{\em Quantum effects}. 
To make an inquiry about the quantum correlations at zero temperature, we use the density matrix renormalization group (DMRG) technique based on a matrix product state representation to evaluate the spin correlations functions~\cite{dolfi2014matrix}. In our calculations, we adopt  $S=3/2$ (which is the spin of  Mn$^{+4}$) and the lattices with $4\times L^2$ sites with $L=4$. The spin-spin correlations normalized by $S^2$  (shown in the bottom panel of Fig.~\ref{fig:correlation}), confirms the transition from N$_{1}$ to N$_{2}$ states at $J_{2c}/J_1\sim 0.1$,  despite the weakening of the correlations as the effect of quantum fluctuations. This is while, the in-plane spins are in the N{\'e}el state, independent of the value of $J_{2c}$ (Fig.~\ref{fig:dmrg}). 

{ Now, we proceed to determine the order of  N$_{1}$-N$_{2}$ transition for the set of exchange interactions obtained by using $U=1.5$ eV. Using Feynman-Hellmann theorem, the first derivative of the ground state energy with respect to a control parameter, say $J_{2c}$, is given by 

\begin{equation}\label{FH}
\frac{\partial E_0}{\partial J_{2c}}=\langle \frac{\partial H}{\partial J_{2c}}\rangle=\frac{3}{2}N 
\langle {\mathbf S}_{i}\cdot {\mathbf S}_{j} \rangle_{2c},
\end{equation}
in which $N=4L^2$ is the number of lattice points in a bilayer honeycomb of linear size $L$ and $\langle {\bf S}_{i}\cdot {\bf S}_{j}\rangle_{2c}$ denotes the spin-spin correlation between the second interlayer neighbors. Figure~\ref{fig:FH} represents the DMRG results for the  variation of  this correlation versus $J_{2c}/J_{1}$, for the lattices of linear sizes $L=2,3,4$. This figure shows a discontinuity in the second interlayer neighbor  spin-spin correlation which becomes more pronounced by increasing the size of the system. Therefore equation~\ref{FH} implies that the phase transition between these two ordered states is first order. Indeed, this result was expected because of different  symmetries of N$_{1}$ and N$_{2}$ states. }

{\em Conclusion}.
In summary, we employed an ab initio LDA+U method to obtain the exchange coupling constants of a spin Hamiltonian for describing the magnetic properties of the honeycomb bilayer BMNO and figure out the reason that this compound does not show any ordering down to very low temperatures. Using $U=1.5$ eV, we found that a Hamiltonian containing only bilinear Heisenberg terms up to third in-plane and fourth out of plane  neighbor, well matches the measured DC magnetic susceptibility for this material. 
Classical MC  simulations and DMRG calculations on this spin Hamiltonian shows
no sign of long-range ordering down to zero temperature. 
%We found that the coupling having the major effect on the frustration of the Hamiltonian is the second inter-plane exchange $J_{2c}$.
It is surprising  that in BMNO the interlayer couplings are tuned in such a way to let this compound living  at the phase boundary of the two collinear magnetic configurations  N$_{1}$ and N$_{2}$. {Indeed, the
interlayer coupling $J_{1c}$ and $J_{3c}$ encourage the N$_{1}$ ordering, while $J_{2c}$
and $J_{4c}$ favor the N$_{2}$ state. Therefore, the balance between these
two sets of couplings adjusts BMNO to be at the N$_{1}$-N$_{2}$ phase
boundary. Hence, in the presence of any imbalance created as the effect of 
tension, compressive pressure, chemical doping, etc, the
transition to N$_{1}$ or N$_{2}$ ordered states is expected.} 
%Hence, we predict that a small change in the inter-layer exchanges, possibly by applying  positive (negative) pressure  normal to the bilayer planes,  will induce a long-range magnetic ordering of type N$_1$ (N$_2$) in this system.} 
 At this very special point, the spin-spin correlations in each layer are N{\'e}el type, however, there is almost a vanishing correlation between the two layers, making the dynamics of the two N{\'e}el states uncorrelated. The lack of correlations between the adjacent layer makes BMNO an effectively two-dimensional Heisenberg system for which there would be no finite temperature phase transition, according to the Mermin-Wagner theorem. 
{It is also worthy to note that the presence of a strong enough spin-lattice interaction,
could induce a  spin-peierls lattice distortion and hence resolve
the spin frustration. However, the reason that such a transition has
not been observed experimentally, could be due to the small spin-orbit
interaction in Mn atom which makes  the temperature scale corresponding to the  magneto-elastic interaction too small to cause an observable static lattice distortion in BMNO down to $50$ mK.}

%We also speculate that the quantum ground state of BMNO is a linear superposition of the two types of N{\'e}el configurations N$_{1}$  and N$_{2}$ which could give rise to a quantum spin liquid state for this compound. Verification of this speculation requires further investigations. 

%whose fine tuning around a critical value ($J_{2c}/J_1\approx 0.105$),  causes a transition  from an antiparallel N{\'e}el ordering of the two honeycomb layers (N$_{1}$) to a parallel N{\'e}el ordering (N$_{2}$). Surprisingly, the value of this coupling calculated by LDA+U, i.e $J_{2c}/J_1=1.02$,  resides in the vicinity of this transition point. 

%A linear spin wave analysis at zero temperature shows that the quantum fluctuation destabilize both the classical ordered state in a narrow region  of normalized second inter-plane coupling, i.e. $0.105 \lesssim J_{2c}/J_1 \lesssim 0.150 $. Whether  this intermediate phase is a quantum spin liquid or not needs further investigations. 

\begin{acknowledgements}
We acknowledge Michel Gingras, Jeff Rau and Stefano de Gironcoli for the most useful discussions and comments. We also thank 
 Phivos Mavropoulos for providing  us with his MC code. DMRG results were checked by using the ALPS {\bf mps-optim} application.
\end{acknowledgements}
%\appendix 
%\clearpage %Should be added directly above bibliography, if 1 sided
%\cleardoublepage %If 2 sided, thanks to egreg
%\bibliographystyle{aps}

%\newpage
\bibliographystyle{apsrev4-1}

\bibliography{bibliography}

%\bibitem{supplement} See Supplementary Material at [URL].

%\end{thebibliography}

\section*{Supplemental  Material}

\section{charge analysis }
\label{app.1}
\begin{figure}[h]
   %\vspace{0.7 cm}
   \includegraphics[width=9cm,angle=0]{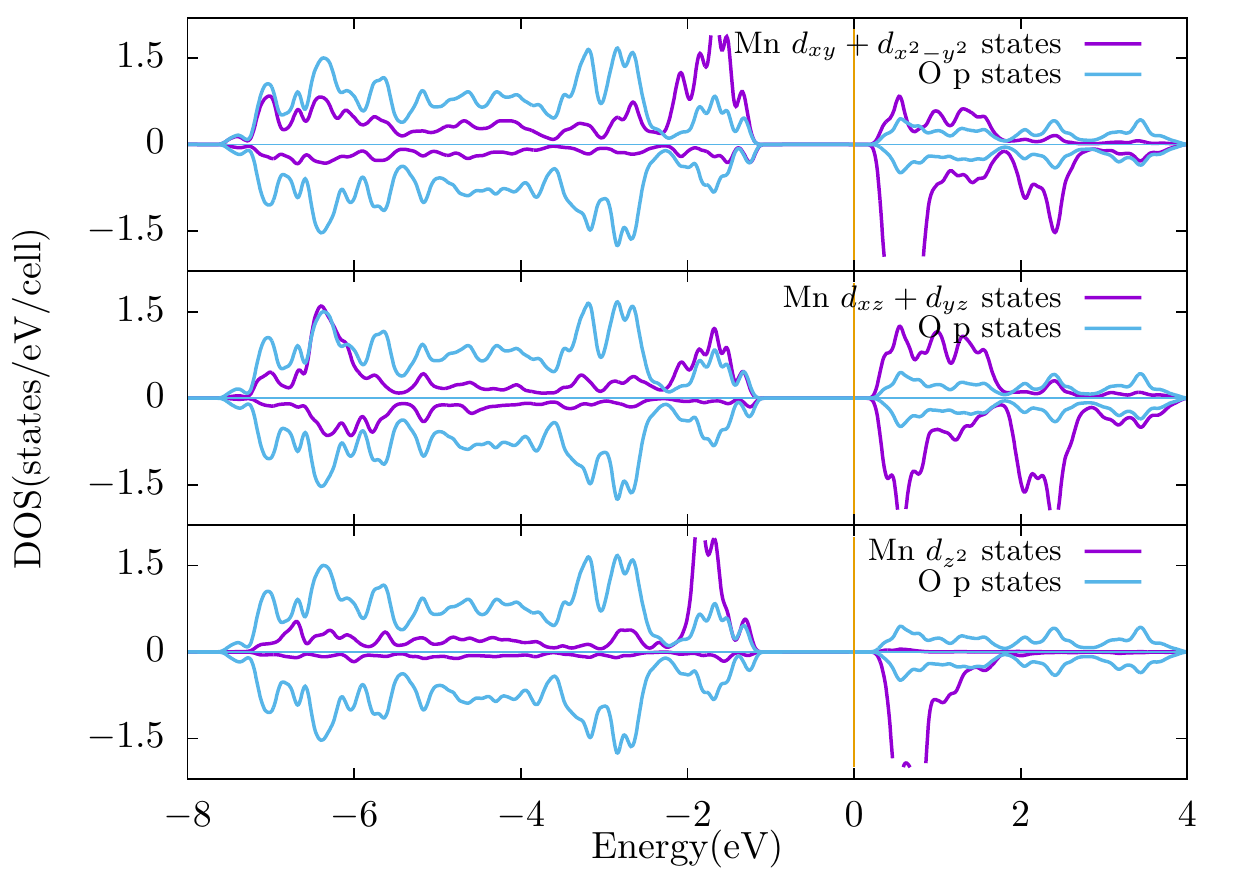}
    \caption{Projected Density of States (PDOS) of Mn-$d$ and O-$p$ orbitals.}
    \label{fig:pdos}
\end{figure}

In this section, we explain why Mn-O bonds has ionic-covalent character. 
In fact, the ionic-covalent character of  Mn-O bonds can be obsereved in
the hybridization of Mn-$d$ and O-$p$ orbitals.
The point group of MnO$_6$ clusters in \bmno is $C_3$, where the $C_3$-axis
 is perpendicular to honeycomb layer. Therefore, the Mn-$d$ orbitals are splitted as the effect of the crystal field  
into $d_{z^2}$, $(d_{xz}, d_{yz})$ and $(d_{xy}, d_{x^2-y^2})$. 
The projected density of states plotted in figure~\ref{fig:pdos} together with the orbital occupation calculation shown in Table.~\ref{tab:d-orb}, indicate that the hybridization among  $d$ orbitals of Mn  and $p$ orbitals of O  makes the crystal field states $(d_{xz}, d_{yz})$ and $(d_{xy}, d_{x^2-y^2})$ to have fractional occupations.
%table1!!!!!!!!!!!!!!!!!!!!!!!!!!!!!!!!!!!!!
\begin{table}[h]
\caption{ Charge distribution among Mn-$d$ orbitals obtained by GGA/PAW Lowdin charge analysis.
}
\label{tab:d-orb}
\begin{ruledtabular}
\centering
%\begin{adjustbox}{max width=\textwidth}
\begin{adjustbox}{max width=16cm}
\begin{tabular}{c|cccc}
spin          &  $d_{tot}$   & $d_{z^2}$ & $(d_{xz}, d_{yz})$ & $(d_{xy}, d_{x^2-y^2})$ \\ \hline
$\uparrow$    & $3.8456$     & $0.9825$  &      $0.6296$      &    $0.8020$      \\
$\downarrow$  & $1.2960$     & $0.1890$  &      $0.2914$      &    $0.2621$
\end{tabular}
\end{adjustbox}
\end{ruledtabular}
\end{table}
Therefore the bond valence sums (Bi$^{3+}_3$Mn$^{4+}_4$O$^{2-}_{12}$(NO3)$^-$) ~\cite{bmnostruct}
does not give rise to a true understanding of
the charge distribution and also about magnetization of Mn atoms.

Figure~\ref{fig:pdos} also shows a wide range of PDOS for all occupied $d$-orbitals, which could be a reason that why the 
onsite electron-electron repulsion  parameter $U$ is not very large in this system.  
%!!!!!!!!!!!!!!!!!!!!!!!!!!!FIG3
%!!!!!!!!!!!!!!!!!!!!!!!!!!!FIG3

\section{Exchange constants}
\label{app.3}

In the experimental structure of \bmno~\cite{bmnostruct} (with $P3$ space group),
within the experimental error, vertical positions of Mn atoms in the unit cell are almost equal (Table~\ref{tab:pos}). 
 In the DFT calculations, we found that  Mn$_1$ and Mn$_2$ 
as well as Mn$_3$ and Mn$_4$ do not have {exactly} the same vertical positions, even 
if exact site geometry optimization is performed. However, as Table~\ref{tab:pos} shows, in DFT, the vertical positions of Mn$_1$ and Mn$_2$ and also Mn$_3$ and Mn$_4$ are very close. 
The site geometry optimization has been done within $0.001$ eV/Ang accuracy.
The space group of geometry optimization structure and experimental structure is the same (P3, No 143).  So the space group of Bi3Mn4O14 doesn't change during geometry optimization. 

%table1!!!!!!!!!!!!!!!!!!!!!!!!!!!!!!!!!!!!!
\begin{table*}[h]
\caption{ Experiment and DFT Wycoff position of Mn atoms. The DFT calculation is done with $U=1.5$eV by FPLO method. }
\label{tab:pos}
\begin{ruledtabular}
\centering
%\begin{adjustbox}{max width=\textwidth}
\begin{adjustbox}{max width=16cm}
\begin{tabular}{c|ccc|ccc}
aotm      & & Experiment Wycoff position  & & &  DFT Wycoff position &  \\ \hline
          &   x    &   y   &     z        &    x  &   y   &      z \\
Mn$_1$    &  2/3   &  1/3  &   0.855(5)   &   2/3 &  1/3  &  0.855500 \\
Mn$_2$    &  1/3   &  2/3  &   0.852(6)   &   1/3 &  2/3  &  0.855979 \\
Mn$_3$    &  2/3   &  1/3  &   0.218(5)   &   2/3 &  1/3  &  0.233593  \\
Mn$_4$    &  1/3   &  2/3  &   0.223(6)   &   1/3 &  2/3  &  0.233109 
\end{tabular}
\end{adjustbox}
\end{ruledtabular}
\end{table*}
Since  Mn atoms are not completely identical, we have 
to use more Heisenberg constants. For example, instead of  only one $J_1$  for first 
nearest neighbor interaction we need to consider two: one between $\rm Mn_1$ and $\rm Mn_2$ ($J_1^{1,2})$ 
and the other between  ${\rm Mn}_3$ and ${\rm Mn}_4$ ($J_1^{3,4}$). For the second neighbor couplings  $J_2$ we also  have $J_2^{1,1}$, 
$J_2^{2,2}$, $J_2^{3,3}$ and $J_2^{4,4}$. Similarly, there are variety of couplings for other inter and intra layer exchange interactions (see Table.~\ref{tab:Js_rx}).  

To calculate the couplings of the Heisenberg Hamiltonian, we use $54$ magnetic 
configurations listed in Table.~\ref{tab:confs}. Employing the least square method by considering these $54$  magnetic configurations,
enables us to calculate the magnetic exchanges  the within accuracy of 0.02 meV.

The ab initio results for the exchange couplings are given in Table.~\ref{tab:Js_rx}. 
These results are obtained after performing geometry optimization in a ferromagnetic configuration. 
The differences between the couplings of the same range  are  small ($\sim 0.1$meV), hence their arithmetic mean are reported
in the main paper. 

Using the  experimental structure, the difference between the couplings of the same range are 
significant  (see Table~\ref{tab:Js_norx}). For example,  $J_1^{1,2}$=27.4 meV
$J_1^{3, 4}$=10.8 meV within GGA/FPLO method.
So assuming identical Mn atoms for experimental structure to derive $J_{ij}$ are
completely wrong. 

{ In the main paper we assumed that the interaction between successive double layers could be ignored. 
To verify this assumption we calculate the energy difference of  two magnetic configurations, a ferromagnetic configuration
in which all the spins in all the bilayers are parallel and  an anti-ferromagnetic configuration in which the spins in each bilayer 
are parallel, while they  are antiparallel to the ones in neighboring bilayers. This energy difference gives an estimate of the strength of interaction  between separated bilayers. 
Using FPLO method with  $U=1.5$ eV, we obtain an energy difference (per Mn atom) of $\sim0.3$ meV for these two configurations.
 This value is an order of magnitude less than  the minimum  energy difference (per Mn atom) of $54$ magnetic configurations inside each bilayer with respect to the ferromagnetic reference state,  which is $6.6$ meV (Fig. 3 of main paper). Therefore it would be  safe to neglect the bilayer-bilayer interaction.}

%table1!!!!!!!!!!!!!!!!!!!!!!!!!!!!!!!!!!!!!!!!!!!!!!!!!!!!!!!!!!!!!!!!!!!!!!!!!!!!!!!!!!!!!!!!!!!!!!!!!!!
\begin{table*}[h]
\caption{Heisenberg constants  obtained  by ab intio calculations (LDA+$U$) using FPLO. The structure, which is used in these calculations, is derived from geometry optimization of \bmno in its feromagetic state. 
}
\label{tab:Js_rx}
\begin{ruledtabular}
\begin{tabular}{c|c|cccccccccccccccc}
method & $U$   &  $J_{1}^{1,2}$ & $J_{1}^{3,4}$ & $J_{2}^{1,1}$ & $J_{2}^{2,2}$ & $J_{2}^{3,3}$ & $J_{2}^{4,4}$ & $J_{3}^{(1,2)}$ & $J_{3}^{3,4}$ & $J_{1c}^{1,3}$ & $J_{1c}^{2,4}$ & $J_{2c}^{1,3}$ & $J_{2c}^{2,4}$ &  $J_{3c}^{1,3}$ & $J_{3c}^{2,4}$ & $J_{4c}^{1,3}$ & $J_{4c}^{2,4}$ \\ \hline 
\multirow{3}{*} {FPLO} &  1.5  &    10.8          &    10.7         &   1.0           &  0.8            &  1.0            &  0.8            &  1.2            &  1.2            &  2.9             &  3.0             &   1.1        &        1.1           &       0.5         &   0.6            &       0.9        &  0.9 \\
                       &  2.0  &    9.0           &    9.0          &   0.9           &  0.7            &  0.9            &  0.7            &  1.0            &  1.0            &  2.6             &  2.6             &   1.0            &    0.9           &       0.4         &   0.5            &       0.8        &  0.8  \\ 
                       &  3.0  &    6.6           &    6.6          &   0.7           &   0.6           &  0.7            &  0.6            &  0.8            &  0.8            &  2.0             &  2.1             &   0.7            &    0.7           &       0.3         &   0.4            &       0.6        &  0.6  \\ 
                       &  4.0  &    5.1           &    5.1          &   0.5           &  0.4            &  0.5            &  0.5            &  0.6            &  0.7            &  1.6             &  1.7             &   0.6            &    0.6           &       0.3         &   0.3            &       0.5        &  0.5  \\ 
\end{tabular}
\end{ruledtabular}
\end{table*}
%table1!!!!!!!!!!!!!!!!!!!!!!!!!!!!!!!!!!!!!!!!!!!!!!!!!!!!!!!!!!!!!!!!!!!!!!!!

%table2!!!!!!!!!!!!!!!!!!!!!!!!!!!!!!!!!!!!!!!!!!!!!!!!!!!!!!!!!!!!!!!!!!!!!!!!!!!!!!!!!!!!!!!!!!!!!!!!!!!
\begin{table*}[h]
\caption{Heisenberg constants  obtained  by ab intio calculations (LDA+$U$) using different $U$ and
different methods. The experimental structure is used in these calculations. 
}
\label{tab:Js_norx}
\begin{ruledtabular}
\begin{tabular}{c|c|cccccccccccccccc}
method & $U$   &  $J_{1}^{1,2}$ & $J_{1}^{3,4}$ & $J_{2}^{1,1}$ & $J_{2}^{2,2}$ & $J_{2}^{3,3}$ & $J_{2}^{4,4}$ & $J_{3}^{1,2}$ & $J_{3}^{3,4}$ & $J_{1c}^{1,3}$ & $J_{1c}^{2,4}$ & $J_{2c}^{1,3}$ & $J_{2c}^{2,4}$ &  $J_{3c}^{1,3}$ & $J_{3c}^{2,4}$ & $J_{4c}^{1,3}$ & $J_{4c}^{2,4}$ \\ \hline 
\multirow{5}{*}{FPLO}  &  0.0  &    27.4          &    10.8         &   0.9           &  0.7            &  2.5            &  2.0            &  1.2            & 2.2             &  4.6             & 5.9              &   1.8            &    1.3           &       0.2         &  0.6             &      1.0          &   0.7   \\
                      &   2.0  &   18.4           &    6.5          &   0.6           &  0.5            & 1.6             &  1.4            &  0.8            & 1.3             &  2.7             & 3.6              &   1.2            &    1.0           &       0.3         &  0.6             &      0.9          &   0.7   \\
                       &  3.0  &   14.1           &    4.3          &   0.5           &  0.4            &  1.2            &  1.1            &  0.6            & 1.0             &  2.2             & 2.8              &   0.9            &    0.8           &       0.3         &  0.4             &      0.7          &   0.5 \\ 
                       &  4.0  &   11.1           &    2.8          &   0.3           &  0.3            &  1.0            &   0.8           &  0.5            &  0.8            &   1.7            &  2.3             &   0.7            &    0.6           &       0.2         &   0.3            &       0.5         &   0.4  \\
\end{tabular}
\end{ruledtabular}
\end{table*}
%table2!!!!!!!!!!!!!!!!!!!!!!!!!!!!!!!!!!!!!!!!!!!!!!!!!!!!!!!!!!!!!!!!!!!!!!!!

\section{Magnetic Configurations}
\label{app.4}
\begin{figure}[h]
   %\vspace{0.7 cm}
   \includegraphics[width=6cm,angle=0]{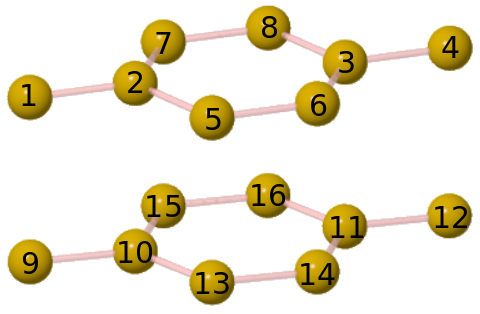}
    \caption{Each Mn atoms are labeled by a number in $2 \times 2 \times 1$ supercell of \bmno}
    \label{fig:labels}
\end{figure}

In this section, we show the $54$ magnetic configurations used for the calculations of the Heisenberg coupling constants.  To represent magnetic configurations, we assign a number  on each Mn atoms in the $2\times 2 \times 1$
supercell of \bmno (figure \ref{fig:labels}), and then we specify the direction of Mn magnetic moments (up or down) by arrows shown in Table.~\ref{tab:confs}.

%We label each
%magnetic configuration by letters. Letters A, B, C, D and E represent magnetic configurations in Ref ~\cite{kandpalprb}.
%!!!!!!!!!!!!!!!!!!!!!!!!!!!!!!!!!!!!!!!!!
%%%%%%%%%%%%%%%%%%%%%%%%%%%%
\begin{table*}[h]
\caption{ $54$ magnetic configurations which are used to derive $J_{ij}$.
}
\label{tab:confs}
\begin{ruledtabular}
\begin{tabular}{c|cccccccccccccccc}
Magnetic configuration &     Mn$_1$      &       Mn$_2$    &      Mn$_3$    &     Mn$_4$      &    Mn$_5$   &    Mn$_6$       &    Mn$_7$       &    Mn$_8$       &    Mn$_9$       &   Mn$_{10}$      &Mn$_{11}$      & Mn$_{12}$     & Mn$_{13}$    &   Mn$_{14}$        &   Mn$_{15}$   & Mn$_{16}$  \\ \hline 
1&  $\uparrow$ &$\uparrow$ &$\uparrow$ &$\uparrow$ &$\uparrow$ &$\uparrow$ &$\uparrow$ &$\uparrow$ &$\uparrow$ &$\uparrow$ &$\uparrow$ &$\uparrow$ &$\uparrow$ &$\uparrow$ &$\uparrow$ &$\uparrow$ \\ 
2&  $\uparrow$ &$\uparrow$ &$\downarrow$ &$\downarrow$ &$\uparrow$ &$\downarrow$ &$\uparrow$ &$\downarrow$ &$\uparrow$ &$\uparrow$ &$\downarrow$ &$\downarrow$ &$\uparrow$ &$\downarrow$ &$\uparrow$ &$\downarrow$ \\ 
3&  $\uparrow$ &$\downarrow$ &$\downarrow$ &$\uparrow$ &$\uparrow$ &$\uparrow$ &$\uparrow$ &$\uparrow$ &$\uparrow$ &$\downarrow$ &$\downarrow$ &$\uparrow$ &$\uparrow$ &$\uparrow$ &$\uparrow$ &$\uparrow$ \\ 
4&  $\uparrow$ &$\downarrow$ &$\uparrow$ &$\uparrow$ &$\uparrow$ &$\uparrow$ &$\uparrow$ &$\uparrow$ &$\uparrow$ &$\uparrow$ &$\uparrow$ &$\uparrow$ &$\uparrow$ &$\uparrow$ &$\uparrow$ &$\uparrow$ \\ 
5&  $\downarrow$ &$\downarrow$ &$\uparrow$ &$\uparrow$ &$\uparrow$ &$\uparrow$ &$\uparrow$ &$\uparrow$ &$\downarrow$ &$\downarrow$ &$\uparrow$ &$\uparrow$ &$\uparrow$ &$\uparrow$ &$\uparrow$ &$\uparrow$ \\ 
6&  $\uparrow$ &$\uparrow$ &$\downarrow$ &$\downarrow$ &$\uparrow$ &$\downarrow$ &$\uparrow$ &$\downarrow$ &$\uparrow$ &$\uparrow$ &$\uparrow$ &$\uparrow$ &$\uparrow$ &$\uparrow$ &$\uparrow$ &$\uparrow$ \\ 
7&  $\uparrow$ &$\uparrow$ &$\downarrow$ &$\downarrow$ &$\downarrow$ &$\downarrow$ &$\downarrow$ &$\downarrow$ &$\uparrow$ &$\uparrow$ &$\uparrow$ &$\uparrow$ &$\uparrow$ &$\uparrow$ &$\uparrow$ &$\uparrow$ \\ 
8&  $\uparrow$ &$\downarrow$ &$\uparrow$ &$\uparrow$ &$\downarrow$ &$\uparrow$ &$\downarrow$ &$\uparrow$ &$\uparrow$ &$\uparrow$ &$\uparrow$ &$\uparrow$ &$\uparrow$ &$\uparrow$ &$\uparrow$ &$\uparrow$ \\ 
9&  $\uparrow$ &$\uparrow$ &$\uparrow$ &$\uparrow$ &$\downarrow$ &$\uparrow$ &$\downarrow$ &$\uparrow$ &$\uparrow$ &$\uparrow$ &$\uparrow$ &$\uparrow$ &$\uparrow$ &$\uparrow$ &$\uparrow$ &$\uparrow$ \\ 
10(N$_1$)& $\uparrow$ &$\downarrow$ &$\uparrow$ &$\downarrow$ &$\uparrow$ &$\downarrow$ &$\uparrow$ &$\downarrow$ &$\downarrow$ &$\uparrow$ &$\downarrow$ &$\uparrow$ &$\downarrow$ &$\uparrow$ &$\downarrow$ &$\uparrow$ \\ 
11&  $\downarrow$ &$\downarrow$ &$\uparrow$ &$\uparrow$ &$\downarrow$ &$\downarrow$ &$\uparrow$ &$\uparrow$ &$\downarrow$ &$\downarrow$ &$\uparrow$ &$\uparrow$ &$\uparrow$ &$\uparrow$ &$\uparrow$ &$\uparrow$ \\ 
12&  $\downarrow$ &$\downarrow$ &$\downarrow$ &$\uparrow$ &$\uparrow$ &$\uparrow$ &$\uparrow$ &$\uparrow$ &$\uparrow$ &$\uparrow$ &$\uparrow$ &$\uparrow$ &$\uparrow$ &$\uparrow$ &$\uparrow$ &$\uparrow$ \\ 
13&  $\downarrow$ &$\downarrow$ &$\downarrow$ &$\uparrow$ &$\uparrow$ &$\uparrow$ &$\uparrow$ &$\uparrow$ &$\downarrow$ &$\downarrow$ &$\downarrow$ &$\uparrow$ &$\uparrow$ &$\uparrow$ &$\uparrow$ &$\uparrow$ \\ 
14&  $\downarrow$ &$\downarrow$ &$\downarrow$ &$\downarrow$ &$\uparrow$ &$\uparrow$ &$\uparrow$ &$\uparrow$ &$\uparrow$ &$\uparrow$ &$\uparrow$ &$\uparrow$ &$\uparrow$ &$\uparrow$ &$\uparrow$ &$\uparrow$ \\ 
15&  $\uparrow$ &$\downarrow$ &$\uparrow$ &$\downarrow$ &$\uparrow$ &$\downarrow$ &$\uparrow$ &$\downarrow$ &$\uparrow$ &$\uparrow$ &$\uparrow$ &$\uparrow$ &$\uparrow$ &$\uparrow$ &$\uparrow$ &$\uparrow$ \\ 
16&  $\uparrow$ &$\downarrow$ &$\uparrow$ &$\downarrow$ &$\uparrow$ &$\downarrow$ &$\uparrow$ &$\downarrow$ &$\uparrow$ &$\downarrow$ &$\uparrow$ &$\uparrow$ &$\uparrow$ &$\uparrow$ &$\uparrow$ &$\uparrow$ \\ 
17&  $\downarrow$ &$\downarrow$ &$\downarrow$ &$\downarrow$ &$\uparrow$ &$\uparrow$ &$\uparrow$ &$\uparrow$ &$\downarrow$ &$\downarrow$ &$\uparrow$ &$\uparrow$ &$\uparrow$ &$\uparrow$ &$\uparrow$ &$\uparrow$ \\ 
18&  $\downarrow$ &$\downarrow$ &$\downarrow$ &$\downarrow$ &$\uparrow$ &$\uparrow$ &$\uparrow$ &$\uparrow$ &$\downarrow$ &$\downarrow$ &$\downarrow$ &$\downarrow$ &$\uparrow$ &$\uparrow$ &$\uparrow$ &$\uparrow$ \\ 
19&  $\uparrow$ &$\downarrow$ &$\uparrow$ &$\downarrow$ &$\uparrow$ &$\downarrow$ &$\uparrow$ &$\downarrow$ &$\uparrow$ &$\downarrow$ &$\downarrow$ &$\uparrow$ &$\uparrow$ &$\uparrow$ &$\uparrow$ &$\uparrow$ \\ 
20&  $\uparrow$ &$\downarrow$ &$\uparrow$ &$\downarrow$ &$\uparrow$ &$\downarrow$ &$\uparrow$ &$\downarrow$ &$\uparrow$ &$\downarrow$ &$\downarrow$ &$\downarrow$ &$\uparrow$ &$\uparrow$ &$\uparrow$ &$\uparrow$ \\ 
21&  $\uparrow$ &$\downarrow$ &$\uparrow$ &$\uparrow$ &$\downarrow$ &$\uparrow$ &$\uparrow$ &$\downarrow$ &$\uparrow$ &$\downarrow$ &$\uparrow$ &$\uparrow$ &$\downarrow$ &$\uparrow$ &$\uparrow$ &$\uparrow$ \\ 
22&  $\downarrow$ &$\downarrow$ &$\downarrow$ &$\downarrow$ &$\uparrow$ &$\uparrow$ &$\uparrow$ &$\uparrow$ &$\uparrow$ &$\uparrow$ &$\downarrow$ &$\uparrow$ &$\downarrow$ &$\uparrow$ &$\uparrow$ &$\uparrow$ \\ 
23&  $\uparrow$ &$\uparrow$ &$\downarrow$ &$\downarrow$ &$\uparrow$ &$\uparrow$ &$\downarrow$ &$\uparrow$ &$\downarrow$ &$\uparrow$ &$\uparrow$ &$\downarrow$ &$\uparrow$ &$\uparrow$ &$\uparrow$ &$\downarrow$ \\ 
24&  $\uparrow$ &$\uparrow$ &$\downarrow$ &$\uparrow$ &$\uparrow$ &$\uparrow$ &$\downarrow$ &$\uparrow$ &$\downarrow$ &$\uparrow$ &$\uparrow$ &$\downarrow$ &$\uparrow$ &$\uparrow$ &$\uparrow$ &$\downarrow$ \\ 
25&  $\downarrow$ &$\uparrow$ &$\downarrow$ &$\downarrow$ &$\downarrow$ &$\downarrow$ &$\downarrow$ &$\downarrow$ &$\downarrow$ &$\uparrow$ &$\downarrow$ &$\uparrow$ &$\uparrow$ &$\uparrow$ &$\downarrow$ &$\uparrow$ \\ 
26&  $\uparrow$ &$\uparrow$ &$\downarrow$ &$\uparrow$ &$\uparrow$ &$\downarrow$ &$\downarrow$ &$\uparrow$ &$\uparrow$ &$\uparrow$ &$\uparrow$ &$\uparrow$ &$\uparrow$ &$\uparrow$ &$\downarrow$ &$\uparrow$ \\ 
27& $\uparrow$ &$\uparrow$ &$\downarrow$ &$\downarrow$ &$\uparrow$ &$\downarrow$ &$\downarrow$ &$\uparrow$ &$\uparrow$ &$\uparrow$ &$\downarrow$ &$\uparrow$ &$\downarrow$ &$\uparrow$ &$\uparrow$ &$\downarrow$ \\ 
28(N$_2$)&  $\uparrow$ &$\downarrow$ &$\uparrow$ &$\downarrow$ &$\uparrow$ &$\downarrow$ &$\uparrow$ &$\downarrow$ &$\uparrow$ &$\downarrow$ &$\uparrow$ &$\downarrow$ &$\uparrow$ &$\downarrow$ &$\uparrow$ &$\downarrow$ \\ 
29& $\downarrow$ &$\downarrow$ &$\downarrow$ &$\downarrow$ &$\uparrow$ &$\downarrow$ &$\uparrow$ &$\downarrow$ &$\uparrow$ &$\uparrow$ &$\downarrow$ &$\uparrow$ &$\downarrow$ &$\uparrow$ &$\uparrow$ &$\uparrow$ \\ 
30& $\downarrow$ &$\downarrow$ &$\downarrow$ &$\uparrow$ &$\downarrow$ &$\uparrow$ &$\downarrow$ &$\uparrow$ &$\downarrow$ &$\downarrow$ &$\uparrow$ &$\downarrow$ &$\downarrow$ &$\uparrow$ &$\uparrow$ &$\uparrow$ \\ 
31& $\downarrow$ &$\uparrow$ &$\downarrow$ &$\uparrow$ &$\uparrow$ &$\uparrow$ &$\uparrow$ &$\uparrow$ &$\uparrow$ &$\downarrow$ &$\uparrow$ &$\uparrow$ &$\downarrow$ &$\downarrow$ &$\downarrow$ &$\uparrow$ \\ 
32& $\uparrow$ &$\uparrow$ &$\uparrow$ &$\uparrow$ &$\downarrow$ &$\downarrow$ &$\downarrow$ &$\uparrow$ &$\uparrow$ &$\uparrow$ &$\downarrow$ &$\uparrow$ &$\downarrow$ &$\uparrow$ &$\downarrow$ &$\uparrow$ \\ 
33& $\uparrow$ &$\uparrow$ &$\uparrow$ &$\downarrow$ &$\uparrow$ &$\downarrow$ &$\downarrow$ &$\downarrow$ &$\downarrow$ &$\uparrow$ &$\downarrow$ &$\downarrow$ &$\downarrow$ &$\uparrow$ &$\uparrow$ &$\uparrow$ \\ 
34& $\downarrow$ &$\downarrow$ &$\uparrow$ &$\downarrow$ &$\uparrow$ &$\uparrow$ &$\downarrow$ &$\downarrow$ &$\downarrow$ &$\downarrow$ &$\downarrow$ &$\uparrow$ &$\uparrow$ &$\downarrow$ &$\uparrow$ &$\uparrow$ \\ 
35& $\downarrow$ &$\uparrow$ &$\downarrow$ &$\downarrow$ &$\downarrow$ &$\uparrow$ &$\downarrow$ &$\downarrow$ &$\downarrow$ &$\downarrow$ &$\uparrow$ &$\uparrow$ &$\downarrow$ &$\downarrow$ &$\downarrow$ &$\downarrow$ \\ 
36& $\downarrow$ &$\uparrow$ &$\downarrow$ &$\downarrow$ &$\uparrow$ &$\uparrow$ &$\downarrow$ &$\downarrow$ &$\uparrow$ &$\uparrow$ &$\uparrow$ &$\uparrow$ &$\uparrow$ &$\uparrow$ &$\downarrow$ &$\uparrow$ \\ 
37& $\downarrow$ &$\downarrow$ &$\uparrow$ &$\uparrow$ &$\downarrow$ &$\downarrow$ &$\uparrow$ &$\downarrow$ &$\downarrow$ &$\uparrow$ &$\uparrow$ &$\uparrow$ &$\downarrow$ &$\uparrow$ &$\downarrow$ &$\downarrow$ \\ 
38& $\uparrow$ &$\uparrow$ &$\uparrow$ &$\downarrow$ &$\uparrow$ &$\uparrow$ &$\uparrow$ &$\downarrow$ &$\uparrow$ &$\uparrow$ &$\downarrow$ &$\downarrow$ &$\downarrow$ &$\uparrow$ &$\uparrow$ &$\uparrow$ \\ 
39& $\uparrow$ &$\downarrow$ &$\downarrow$ &$\uparrow$ &$\downarrow$ &$\downarrow$ &$\uparrow$ &$\downarrow$ &$\downarrow$ &$\downarrow$ &$\downarrow$ &$\uparrow$ &$\uparrow$ &$\uparrow$ &$\downarrow$ &$\downarrow$ \\ 
40& $\uparrow$ &$\downarrow$ &$\uparrow$ &$\uparrow$ &$\uparrow$ &$\uparrow$ &$\uparrow$ &$\uparrow$ &$\uparrow$ &$\downarrow$ &$\downarrow$ &$\downarrow$ &$\downarrow$ &$\uparrow$ &$\uparrow$ &$\uparrow$ \\ 
41& $\downarrow$ &$\uparrow$ &$\uparrow$ &$\uparrow$ &$\uparrow$ &$\downarrow$ &$\downarrow$ &$\downarrow$ &$\uparrow$ &$\downarrow$ &$\uparrow$ &$\downarrow$ &$\downarrow$ &$\downarrow$ &$\uparrow$ &$\uparrow$ \\ 
42& $\uparrow$ &$\uparrow$ &$\uparrow$ &$\downarrow$ &$\downarrow$ &$\downarrow$ &$\downarrow$ &$\downarrow$ &$\uparrow$ &$\uparrow$ &$\uparrow$ &$\uparrow$ &$\uparrow$ &$\downarrow$ &$\uparrow$ &$\downarrow$ \\ 
43& $\uparrow$ &$\uparrow$ &$\downarrow$ &$\downarrow$ &$\downarrow$ &$\downarrow$ &$\downarrow$ &$\uparrow$ &$\downarrow$ &$\downarrow$ &$\downarrow$ &$\downarrow$ &$\downarrow$ &$\uparrow$ &$\uparrow$ &$\uparrow$ \\ 
44& $\downarrow$ &$\uparrow$ &$\uparrow$ &$\uparrow$ &$\uparrow$ &$\uparrow$ &$\uparrow$ &$\uparrow$ &$\downarrow$ &$\downarrow$ &$\downarrow$ &$\uparrow$ &$\uparrow$ &$\uparrow$ &$\uparrow$ &$\downarrow$ \\ 
45& $\downarrow$ &$\uparrow$ &$\uparrow$ &$\downarrow$ &$\downarrow$ &$\uparrow$ &$\downarrow$ &$\downarrow$ &$\downarrow$ &$\downarrow$ &$\downarrow$ &$\downarrow$ &$\uparrow$ &$\uparrow$ &$\uparrow$ &$\uparrow$ \\ 
46& $\uparrow$ &$\uparrow$ &$\downarrow$ &$\downarrow$ &$\downarrow$ &$\uparrow$ &$\downarrow$ &$\uparrow$ &$\uparrow$ &$\uparrow$ &$\uparrow$ &$\downarrow$ &$\uparrow$ &$\downarrow$ &$\downarrow$ &$\uparrow$ \\ 
47(N$_3)$& $\uparrow$ &$\uparrow$ &$\uparrow$ &$\uparrow$ &$\uparrow$ &$\uparrow$ &$\uparrow$ &$\uparrow$ &$\downarrow$ &$\downarrow$ &$\downarrow$ &$\downarrow$ &$\downarrow$ &$\downarrow$ &$\downarrow$ &$\downarrow$ \\ 
48& $\downarrow$ &$\downarrow$ &$\downarrow$ &$\downarrow$ &$\downarrow$ &$\downarrow$ &$\downarrow$ &$\uparrow$ &$\uparrow$ &$\uparrow$ &$\uparrow$ &$\uparrow$ &$\downarrow$ &$\downarrow$ &$\downarrow$ &$\uparrow$ \\ 
49& $\uparrow$ &$\downarrow$ &$\downarrow$ &$\downarrow$ &$\downarrow$ &$\uparrow$ &$\downarrow$ &$\downarrow$ &$\downarrow$ &$\downarrow$ &$\uparrow$ &$\downarrow$ &$\downarrow$ &$\downarrow$ &$\downarrow$ &$\uparrow$ \\ 
50& $\downarrow$ &$\downarrow$ &$\uparrow$ &$\downarrow$ &$\uparrow$ &$\uparrow$ &$\uparrow$ &$\downarrow$ &$\downarrow$ &$\downarrow$ &$\downarrow$ &$\uparrow$ &$\downarrow$ &$\uparrow$ &$\uparrow$ &$\downarrow$ \\ 
51& $\uparrow$ &$\uparrow$ &$\downarrow$ &$\uparrow$ &$\uparrow$ &$\downarrow$ &$\uparrow$ &$\downarrow$ &$\downarrow$ &$\uparrow$ &$\uparrow$ &$\downarrow$ &$\downarrow$ &$\uparrow$ &$\downarrow$ &$\uparrow$ \\ 
52& $\downarrow$ &$\uparrow$ &$\downarrow$ &$\downarrow$ &$\downarrow$ &$\downarrow$ &$\downarrow$ &$\uparrow$ &$\downarrow$ &$\uparrow$ &$\downarrow$ &$\uparrow$ &$\downarrow$ &$\uparrow$ &$\downarrow$ &$\downarrow$ \\ 
53& $\uparrow$ &$\uparrow$ &$\downarrow$ &$\uparrow$ &$\uparrow$ &$\downarrow$ &$\uparrow$ &$\uparrow$ &$\downarrow$ &$\downarrow$ &$\uparrow$ &$\uparrow$ &$\downarrow$ &$\downarrow$ &$\downarrow$ &$\uparrow$ \\ 
54& $\downarrow$ &$\uparrow$ &$\downarrow$ &$\uparrow$ &$\downarrow$ &$\uparrow$ &$\uparrow$ &$\uparrow$ &$\downarrow$ &$\downarrow$ &$\uparrow$ &$\uparrow$ &$\downarrow$ &$\uparrow$ &$\downarrow$ &$\uparrow$ \\ 

\end{tabular}
\end{ruledtabular}
\end{table*}
%table2!!!!!!!!!!!!!!!!!!!!!!!!!!!!!!!!!!!!!!!!!!!!!!!!!!!!!!!!!!!!!!!!!!!!!!!!

\section{DMRG results analysis}
We use density matrix renormalization group (DMRG) techniques based on a matrix product state (MPS) representation 
to evaluate the spin correlation functions. MPS is an variational ansatz that the variational parameters can be controlled
by the matrix size, M, called the bond dimension. The ground state converges after sweeping a few times through the system.

We carry out 10 sweeps to converge the ground state 
within an error less than $3$ percent near the phase transition point. Moreover, we compare the correlation functions calculated 
using different bond dimension and find that the error of the results is less than $1.3$ percent. Away from the phase transition point,
 the errors are less than these values and reach to $0.4$ percent.

\bibliographystyle{apsrev4-1}
\bibliography{bibliography}

\begin{thebibliography}{23}%
\makeatletter
\providecommand \@ifxundefined [1]{%
 \@ifx{#1\undefined}
}%
\providecommand \@ifnum [1]{%
 \ifnum #1\expandafter \@firstoftwo
 \else \expandafter \@secondoftwo
 \fi
}%
\providecommand \@ifx [1]{%
 \ifx #1\expandafter \@firstoftwo
 \else \expandafter \@secondoftwo
 \fi
}%
\providecommand \natexlab [1]{#1}%
\providecommand \enquote  [1]{``#1''}%
\providecommand \bibnamefont  [1]{#1}%
\providecommand \bibfnamefont [1]{#1}%
\providecommand \citenamefont [1]{#1}%
\providecommand \href@noop [0]{\@secondoftwo}%
\providecommand \href [0]{\begingroup \@sanitize@url \@href}%
\providecommand \@href[1]{\@@startlink{#1}\@@href}%
\providecommand \@@href[1]{\endgroup#1\@@endlink}%
\providecommand \@sanitize@url [0]{\catcode `\\12\catcode `\$12\catcode
  `\&12\catcode `\#12\catcode `\^12\catcode `\_12\catcode `\%12\relax}%
\providecommand \@@startlink[1]{}%
\providecommand \@@endlink[0]{}%
\providecommand \url  [0]{\begingroup\@sanitize@url \@url }%
\providecommand \@url [1]{\endgroup\@href {#1}{\urlprefix }}%
\providecommand \urlprefix  [0]{URL }%
\providecommand \Eprint [0]{\href }%
\providecommand \doibase [0]{http://dx.doi.org/}%
\providecommand \selectlanguage [0]{\@gobble}%
\providecommand \bibinfo  [0]{\@secondoftwo}%
\providecommand \bibfield  [0]{\@secondoftwo}%
\providecommand \translation [1]{[#1]}%
\providecommand \BibitemOpen [0]{}%
\providecommand \bibitemStop [0]{}%
\providecommand \bibitemNoStop [0]{.\EOS\space}%
\providecommand \EOS [0]{\spacefactor3000\relax}%
\providecommand \BibitemShut  [1]{\csname bibitem#1\endcsname}%
\let\auto@bib@innerbib\@empty
%</preamble>
\bibitem [{\citenamefont {Smirnova}\ \emph {et~al.}(2009)\citenamefont
  {Smirnova}, \citenamefont {Azuma}, \citenamefont {Kumada}, \citenamefont
  {Kusano}, \citenamefont {Matsuda}, \citenamefont {Shimakawa}, \citenamefont
  {Takei}, \citenamefont {Yonesaki},\ and\ \citenamefont
  {Kinomura}}]{bmnostruct}%
  \BibitemOpen
  \bibfield  {author} {\bibinfo {author} {\bibfnamefont {O.}~\bibnamefont
  {Smirnova}}, \bibinfo {author} {\bibfnamefont {M.}~\bibnamefont {Azuma}},
  \bibinfo {author} {\bibfnamefont {N.}~\bibnamefont {Kumada}}, \bibinfo
  {author} {\bibfnamefont {Y.}~\bibnamefont {Kusano}}, \bibinfo {author}
  {\bibfnamefont {M.}~\bibnamefont {Matsuda}}, \bibinfo {author} {\bibfnamefont
  {Y.}~\bibnamefont {Shimakawa}}, \bibinfo {author} {\bibfnamefont
  {T.}~\bibnamefont {Takei}}, \bibinfo {author} {\bibfnamefont
  {Y.}~\bibnamefont {Yonesaki}}, \ and\ \bibinfo {author} {\bibfnamefont
  {N.}~\bibnamefont {Kinomura}},\ }\href@noop {} {\bibfield  {journal}
  {\bibinfo  {journal} {Journal of the American Chemical Society}\ }\textbf
  {\bibinfo {volume} {131}},\ \bibinfo {pages} {8313} (\bibinfo {year}
  {2009})}\BibitemShut {NoStop}%
\bibitem [{\citenamefont {Onishi}\ \emph {et~al.}(2012)\citenamefont {Onishi},
  \citenamefont {Oka}, \citenamefont {Azuma}, \citenamefont {Shimakawa},
  \citenamefont {Motome}, \citenamefont {Taniguchi}, \citenamefont {Hiraishi},
  \citenamefont {Miyazaki}, \citenamefont {Masuda}, \citenamefont {Koda} \emph
  {et~al.}}]{masudaprb}%
  \BibitemOpen
  \bibfield  {author} {\bibinfo {author} {\bibfnamefont {N.}~\bibnamefont
  {Onishi}}, \bibinfo {author} {\bibfnamefont {K.}~\bibnamefont {Oka}},
  \bibinfo {author} {\bibfnamefont {M.}~\bibnamefont {Azuma}}, \bibinfo
  {author} {\bibfnamefont {Y.}~\bibnamefont {Shimakawa}}, \bibinfo {author}
  {\bibfnamefont {Y.}~\bibnamefont {Motome}}, \bibinfo {author} {\bibfnamefont
  {T.}~\bibnamefont {Taniguchi}}, \bibinfo {author} {\bibfnamefont
  {M.}~\bibnamefont {Hiraishi}}, \bibinfo {author} {\bibfnamefont
  {M.}~\bibnamefont {Miyazaki}}, \bibinfo {author} {\bibfnamefont
  {T.}~\bibnamefont {Masuda}}, \bibinfo {author} {\bibfnamefont
  {A.}~\bibnamefont {Koda}},  \emph {et~al.},\ }\href@noop {} {\bibfield
  {journal} {\bibinfo  {journal} {Physical Review B}\ }\textbf {\bibinfo
  {volume} {85}},\ \bibinfo {pages} {184412} (\bibinfo {year}
  {2012})}\BibitemShut {NoStop}%
\bibitem [{\citenamefont {Matsuda}\ \emph {et~al.}(2010)\citenamefont
  {Matsuda}, \citenamefont {Azuma}, \citenamefont {Tokunaga}, \citenamefont
  {Shimakawa},\ and\ \citenamefont {Kumada}}]{matsudaprl}%
  \BibitemOpen
  \bibfield  {author} {\bibinfo {author} {\bibfnamefont {M.}~\bibnamefont
  {Matsuda}}, \bibinfo {author} {\bibfnamefont {M.}~\bibnamefont {Azuma}},
  \bibinfo {author} {\bibfnamefont {M.}~\bibnamefont {Tokunaga}}, \bibinfo
  {author} {\bibfnamefont {Y.}~\bibnamefont {Shimakawa}}, \ and\ \bibinfo
  {author} {\bibfnamefont {N.}~\bibnamefont {Kumada}},\ }\href {\doibase
  10.1103/PhysRevLett.105.187201} {\bibfield  {journal} {\bibinfo  {journal}
  {Phys. Rev. Lett.}\ }\textbf {\bibinfo {volume} {105}},\ \bibinfo {pages}
  {187201} (\bibinfo {year} {2010})}\BibitemShut {NoStop}%
\bibitem [{\citenamefont {Okubo}\ \emph {et~al.}(2010)\citenamefont {Okubo},
  \citenamefont {Elmasry}, \citenamefont {Zhang}, \citenamefont {Fujisawa},
  \citenamefont {Sakurai}, \citenamefont {Ohta}, \citenamefont {Azuma},
  \citenamefont {Sumirnova},\ and\ \citenamefont {Kumada}}]{ESR}%
  \BibitemOpen
  \bibfield  {author} {\bibinfo {author} {\bibfnamefont {S.}~\bibnamefont
  {Okubo}}, \bibinfo {author} {\bibfnamefont {F.}~\bibnamefont {Elmasry}},
  \bibinfo {author} {\bibfnamefont {W.}~\bibnamefont {Zhang}}, \bibinfo
  {author} {\bibfnamefont {M.}~\bibnamefont {Fujisawa}}, \bibinfo {author}
  {\bibfnamefont {T.}~\bibnamefont {Sakurai}}, \bibinfo {author} {\bibfnamefont
  {H.}~\bibnamefont {Ohta}}, \bibinfo {author} {\bibfnamefont {M.}~\bibnamefont
  {Azuma}}, \bibinfo {author} {\bibfnamefont {O.~A.}\ \bibnamefont
  {Sumirnova}}, \ and\ \bibinfo {author} {\bibfnamefont {N.}~\bibnamefont
  {Kumada}},\ }\href {http://stacks.iop.org/1742-6596/200/i=2/a=022042}
  {\bibfield  {journal} {\bibinfo  {journal} {Journal of Physics: Conference
  Series}\ }\textbf {\bibinfo {volume} {200}},\ \bibinfo {pages} {022042}
  (\bibinfo {year} {2010})}\BibitemShut {NoStop}%
\bibitem [{\citenamefont {Kandpal}\ and\ \citenamefont {van~den
  Brink}(2011)}]{kandpalprb}%
  \BibitemOpen
  \bibfield  {author} {\bibinfo {author} {\bibfnamefont {H.~C.}\ \bibnamefont
  {Kandpal}}\ and\ \bibinfo {author} {\bibfnamefont {J.}~\bibnamefont {van~den
  Brink}},\ }\href@noop {} {\bibfield  {journal} {\bibinfo  {journal} {Physical
  Review B}\ }\textbf {\bibinfo {volume} {83}},\ \bibinfo {pages} {140412}
  (\bibinfo {year} {2011})}\BibitemShut {NoStop}%
\bibitem [{\citenamefont {Ganesh}\ \emph
  {et~al.}(2011{\natexlab{a}})\citenamefont {Ganesh}, \citenamefont {Sheng},
  \citenamefont {Kim},\ and\ \citenamefont {Paramekanti}}]{ganesh2011quantum}%
  \BibitemOpen
  \bibfield  {author} {\bibinfo {author} {\bibfnamefont {R.}~\bibnamefont
  {Ganesh}}, \bibinfo {author} {\bibfnamefont {D.}~\bibnamefont {Sheng}},
  \bibinfo {author} {\bibfnamefont {Y.-J.}\ \bibnamefont {Kim}}, \ and\
  \bibinfo {author} {\bibfnamefont {A.}~\bibnamefont {Paramekanti}},\
  }\href@noop {} {\bibfield  {journal} {\bibinfo  {journal} {Physical Review
  B}\ }\textbf {\bibinfo {volume} {83}},\ \bibinfo {pages} {144414} (\bibinfo
  {year} {2011}{\natexlab{a}})}\BibitemShut {NoStop}%
\bibitem [{\citenamefont {Ganesh}\ \emph
  {et~al.}(2011{\natexlab{b}})\citenamefont {Ganesh}, \citenamefont {Isakov},\
  and\ \citenamefont {Paramekanti}}]{ganesh2011neel}%
  \BibitemOpen
  \bibfield  {author} {\bibinfo {author} {\bibfnamefont {R.}~\bibnamefont
  {Ganesh}}, \bibinfo {author} {\bibfnamefont {S.~V.}\ \bibnamefont {Isakov}},
  \ and\ \bibinfo {author} {\bibfnamefont {A.}~\bibnamefont {Paramekanti}},\
  }\href@noop {} {\bibfield  {journal} {\bibinfo  {journal} {Physical Review
  B}\ }\textbf {\bibinfo {volume} {84}},\ \bibinfo {pages} {214412} (\bibinfo
  {year} {2011}{\natexlab{b}})}\BibitemShut {NoStop}%
\bibitem [{\citenamefont {Oitmaa}\ and\ \citenamefont
  {Singh}(2012)}]{oitmaa2012ground}%
  \BibitemOpen
  \bibfield  {author} {\bibinfo {author} {\bibfnamefont {J.}~\bibnamefont
  {Oitmaa}}\ and\ \bibinfo {author} {\bibfnamefont {R.}~\bibnamefont {Singh}},\
  }\href@noop {} {\bibfield  {journal} {\bibinfo  {journal} {Physical Review
  B}\ }\textbf {\bibinfo {volume} {85}},\ \bibinfo {pages} {014428} (\bibinfo
  {year} {2012})}\BibitemShut {NoStop}%
\bibitem [{\citenamefont {Okubo}\ \emph {et~al.}(2012)\citenamefont {Okubo},
  \citenamefont {Ueda}, \citenamefont {Ohta}, \citenamefont {Zhang},
  \citenamefont {Sakurai}, \citenamefont {Onishi}, \citenamefont {Azuma},
  \citenamefont {Shimakawa}, \citenamefont {Nakano},\ and\ \citenamefont
  {Sakai}}]{dmbmno}%
  \BibitemOpen
  \bibfield  {author} {\bibinfo {author} {\bibfnamefont {S.}~\bibnamefont
  {Okubo}}, \bibinfo {author} {\bibfnamefont {T.}~\bibnamefont {Ueda}},
  \bibinfo {author} {\bibfnamefont {H.}~\bibnamefont {Ohta}}, \bibinfo {author}
  {\bibfnamefont {W.}~\bibnamefont {Zhang}}, \bibinfo {author} {\bibfnamefont
  {T.}~\bibnamefont {Sakurai}}, \bibinfo {author} {\bibfnamefont
  {N.}~\bibnamefont {Onishi}}, \bibinfo {author} {\bibfnamefont
  {M.}~\bibnamefont {Azuma}}, \bibinfo {author} {\bibfnamefont
  {Y.}~\bibnamefont {Shimakawa}}, \bibinfo {author} {\bibfnamefont
  {H.}~\bibnamefont {Nakano}}, \ and\ \bibinfo {author} {\bibfnamefont
  {T.}~\bibnamefont {Sakai}},\ }\href {\doibase 10.1103/PhysRevB.86.140401}
  {\bibfield  {journal} {\bibinfo  {journal} {Phys. Rev. B}\ }\textbf {\bibinfo
  {volume} {86}},\ \bibinfo {pages} {140401} (\bibinfo {year}
  {2012})}\BibitemShut {NoStop}%
\bibitem [{\citenamefont {Zhang}\ \emph {et~al.}(2014)\citenamefont {Zhang},
  \citenamefont {Arlego},\ and\ \citenamefont {Lamas}}]{zhang2014quantum}%
  \BibitemOpen
  \bibfield  {author} {\bibinfo {author} {\bibfnamefont {H.}~\bibnamefont
  {Zhang}}, \bibinfo {author} {\bibfnamefont {M.}~\bibnamefont {Arlego}}, \
  and\ \bibinfo {author} {\bibfnamefont {C.}~\bibnamefont {Lamas}},\
  }\href@noop {} {\bibfield  {journal} {\bibinfo  {journal} {Physical Review
  B}\ }\textbf {\bibinfo {volume} {89}},\ \bibinfo {pages} {024403} (\bibinfo
  {year} {2014})}\BibitemShut {NoStop}%
\bibitem [{\citenamefont {Albarrac{\'\i}n}\ and\ \citenamefont
  {Rosales}(2016)}]{albarracin2016field}%
  \BibitemOpen
  \bibfield  {author} {\bibinfo {author} {\bibfnamefont {F.~G.}\ \bibnamefont
  {Albarrac{\'\i}n}}\ and\ \bibinfo {author} {\bibfnamefont {H.}~\bibnamefont
  {Rosales}},\ }\href@noop {} {\bibfield  {journal} {\bibinfo  {journal}
  {Physical Review B}\ }\textbf {\bibinfo {volume} {93}},\ \bibinfo {pages}
  {144413} (\bibinfo {year} {2016})}\BibitemShut {NoStop}%
\bibitem [{\citenamefont {Bishop}\ and\ \citenamefont
  {Li}(2016)}]{bishop2016frustrated}%
  \BibitemOpen
  \bibfield  {author} {\bibinfo {author} {\bibfnamefont {R.}~\bibnamefont
  {Bishop}}\ and\ \bibinfo {author} {\bibfnamefont {P.}~\bibnamefont {Li}},\
  }\href@noop {} {\bibfield  {journal} {\bibinfo  {journal} {arXiv preprint
  arXiv:1611.03287}\ } (\bibinfo {year} {2016})}\BibitemShut {NoStop}%
\bibitem [{\citenamefont {Koepernik}\ and\ \citenamefont
  {Eschrig}(1999)}]{FPLO}%
  \BibitemOpen
  \bibfield  {author} {\bibinfo {author} {\bibfnamefont {K.}~\bibnamefont
  {Koepernik}}\ and\ \bibinfo {author} {\bibfnamefont {H.}~\bibnamefont
  {Eschrig}},\ }\href {\doibase 10.1103/PhysRevB.59.1743} {\bibfield  {journal}
  {\bibinfo  {journal} {Phys. Rev. B}\ }\textbf {\bibinfo {volume} {59}},\
  \bibinfo {pages} {1743} (\bibinfo {year} {1999})}\BibitemShut {NoStop}%
\bibitem [{\citenamefont {Giannozzi}\ \emph {et~al.}(2009)\citenamefont
  {Giannozzi} \emph {et~al.}}]{QE}%
  \BibitemOpen
  \bibfield  {author} {\bibinfo {author} {\bibfnamefont {P.}~\bibnamefont
  {Giannozzi}} \emph {et~al.},\ }\href
  {http://stacks.iop.org/0953-8984/21/i=39/a=395502} {\bibfield  {journal}
  {\bibinfo  {journal} {Journal of Physics: Condensed Matter}\ }\textbf
  {\bibinfo {volume} {21}},\ \bibinfo {pages} {395502} (\bibinfo {year}
  {2009})}\BibitemShut {NoStop}%
\bibitem [{\citenamefont {Perdew}\ \emph {et~al.}(1996)\citenamefont {Perdew},
  \citenamefont {Burke},\ and\ \citenamefont {Ernzerhof}}]{PBE}%
  \BibitemOpen
  \bibfield  {author} {\bibinfo {author} {\bibfnamefont {J.~P.}\ \bibnamefont
  {Perdew}}, \bibinfo {author} {\bibfnamefont {K.}~\bibnamefont {Burke}}, \
  and\ \bibinfo {author} {\bibfnamefont {M.}~\bibnamefont {Ernzerhof}},\ }\href
  {\doibase 10.1103/PhysRevLett.77.3865} {\bibfield  {journal} {\bibinfo
  {journal} {Phys. Rev. Lett.}\ }\textbf {\bibinfo {volume} {77}},\ \bibinfo
  {pages} {3865} (\bibinfo {year} {1996})}\BibitemShut {NoStop}%
\bibitem [{\citenamefont {Anisimov}\ \emph {et~al.}(1991)\citenamefont
  {Anisimov}, \citenamefont {Zaanen},\ and\ \citenamefont
  {Andersen}}]{anisimov1991}%
  \BibitemOpen
  \bibfield  {author} {\bibinfo {author} {\bibfnamefont {V.~I.}\ \bibnamefont
  {Anisimov}}, \bibinfo {author} {\bibfnamefont {J.}~\bibnamefont {Zaanen}}, \
  and\ \bibinfo {author} {\bibfnamefont {O.~K.}\ \bibnamefont {Andersen}},\
  }\href {\doibase 10.1103/PhysRevB.44.943} {\bibfield  {journal} {\bibinfo
  {journal} {Phys. Rev. B}\ }\textbf {\bibinfo {volume} {44}},\ \bibinfo
  {pages} {943} (\bibinfo {year} {1991})}\BibitemShut {NoStop}%
\bibitem [{\citenamefont {Anisimov}\ \emph {et~al.}(1993)\citenamefont
  {Anisimov}, \citenamefont {Solovyev}, \citenamefont {Korotin}, \citenamefont
  {Czy\ifmmode~\dot{z}\else \.{z}\fi{}yk},\ and\ \citenamefont
  {Sawatzky}}]{anisimov1993}%
  \BibitemOpen
  \bibfield  {author} {\bibinfo {author} {\bibfnamefont {V.~I.}\ \bibnamefont
  {Anisimov}}, \bibinfo {author} {\bibfnamefont {I.~V.}\ \bibnamefont
  {Solovyev}}, \bibinfo {author} {\bibfnamefont {M.~A.}\ \bibnamefont
  {Korotin}}, \bibinfo {author} {\bibfnamefont {M.~T.}\ \bibnamefont
  {Czy\ifmmode~\dot{z}\else \.{z}\fi{}yk}}, \ and\ \bibinfo {author}
  {\bibfnamefont {G.~A.}\ \bibnamefont {Sawatzky}},\ }\href {\doibase
  10.1103/PhysRevB.48.16929} {\bibfield  {journal} {\bibinfo  {journal} {Phys.
  Rev. B}\ }\textbf {\bibinfo {volume} {48}},\ \bibinfo {pages} {16929}
  (\bibinfo {year} {1993})}\BibitemShut {NoStop}%
\bibitem [{\citenamefont {Liechtenstein}\ \emph {et~al.}(1995)\citenamefont
  {Liechtenstein}, \citenamefont {Anisimov},\ and\ \citenamefont
  {Zaanen}}]{LDAU}%
  \BibitemOpen
  \bibfield  {author} {\bibinfo {author} {\bibfnamefont {A.~I.}\ \bibnamefont
  {Liechtenstein}}, \bibinfo {author} {\bibfnamefont {V.~I.}\ \bibnamefont
  {Anisimov}}, \ and\ \bibinfo {author} {\bibfnamefont {J.}~\bibnamefont
  {Zaanen}},\ }\href {\doibase 10.1103/PhysRevB.52.R5467} {\bibfield  {journal}
  {\bibinfo  {journal} {Phys. Rev. B}\ }\textbf {\bibinfo {volume} {52}},\
  \bibinfo {pages} {R5467} (\bibinfo {year} {1995})}\BibitemShut {NoStop}%
\bibitem [{\citenamefont {Eschrig}\ \emph {et~al.}(2003)\citenamefont
  {Eschrig}, \citenamefont {Koepernik},\ and\ \citenamefont
  {Chaplygin}}]{FPLOLDAU}%
  \BibitemOpen
  \bibfield  {author} {\bibinfo {author} {\bibfnamefont {H.}~\bibnamefont
  {Eschrig}}, \bibinfo {author} {\bibfnamefont {K.}~\bibnamefont {Koepernik}},
  \ and\ \bibinfo {author} {\bibfnamefont {I.}~\bibnamefont {Chaplygin}},\
  }\href {\doibase http://dx.doi.org/10.1016/S0022-4596(03)00274-3} {\bibfield
  {journal} {\bibinfo  {journal} {Journal of Solid State Chemistry}\ }\textbf
  {\bibinfo {volume} {176}},\ \bibinfo {pages} {482 } (\bibinfo {year}
  {2003})},\ \bibinfo {note} {special issue on The Impact of Theoretical
  Methods on Solid-State Chemistry}\BibitemShut {NoStop}%
\bibitem [{sup()}]{supplement}%
  \BibitemOpen
  \href@noop {} {}\bibinfo {howpublished} {See Supplementary Material at
  [URL].}\BibitemShut {Stop}%
\bibitem [{\citenamefont {de-la Roza}\ \emph {et~al.}(2014)\citenamefont {de-la
  Roza}, \citenamefont {Johnson},\ and\ \citenamefont {Luaña}}]{critic2}%
  \BibitemOpen
  \bibfield  {author} {\bibinfo {author} {\bibfnamefont {A.~O.}\ \bibnamefont
  {de-la Roza}}, \bibinfo {author} {\bibfnamefont {E.~R.}\ \bibnamefont
  {Johnson}}, \ and\ \bibinfo {author} {\bibfnamefont {V.}~\bibnamefont
  {Luaña}},\ }\href {\doibase http://dx.doi.org/10.1016/j.cpc.2013.10.026}
  {\bibfield  {journal} {\bibinfo  {journal} {Computer Physics Communications}\
  }\textbf {\bibinfo {volume} {185}},\ \bibinfo {pages} {1007 } (\bibinfo
  {year} {2014})}\BibitemShut {NoStop}%
\bibitem [{\citenamefont {de-la Roza}\ \emph {et~al.}(2009)\citenamefont {de-la
  Roza}, \citenamefont {Blanco}, \citenamefont {Pendás},\ and\ \citenamefont
  {Luaña}}]{critic}%
  \BibitemOpen
  \bibfield  {author} {\bibinfo {author} {\bibfnamefont {A.~O.}\ \bibnamefont
  {de-la Roza}}, \bibinfo {author} {\bibfnamefont {M.}~\bibnamefont {Blanco}},
  \bibinfo {author} {\bibfnamefont {A.~M.}\ \bibnamefont {Pendás}}, \ and\
  \bibinfo {author} {\bibfnamefont {V.}~\bibnamefont {Luaña}},\ }\href
  {\doibase http://dx.doi.org/10.1016/j.cpc.2008.07.018} {\bibfield  {journal}
  {\bibinfo  {journal} {Computer Physics Communications}\ }\textbf {\bibinfo
  {volume} {180}},\ \bibinfo {pages} {157 } (\bibinfo {year}
  {2009})}\BibitemShut {NoStop}%
\bibitem [{\citenamefont {Dolfi}\ \emph {et~al.}(2014)\citenamefont {Dolfi},
  \citenamefont {Bauer}, \citenamefont {Keller}, \citenamefont {Kosenkov},
  \citenamefont {Ewart}, \citenamefont {Kantian}, \citenamefont {Giamarchi},\
  and\ \citenamefont {Troyer}}]{dolfi2014matrix}%
  \BibitemOpen
  \bibfield  {author} {\bibinfo {author} {\bibfnamefont {M.}~\bibnamefont
  {Dolfi}}, \bibinfo {author} {\bibfnamefont {B.}~\bibnamefont {Bauer}},
  \bibinfo {author} {\bibfnamefont {S.}~\bibnamefont {Keller}}, \bibinfo
  {author} {\bibfnamefont {A.}~\bibnamefont {Kosenkov}}, \bibinfo {author}
  {\bibfnamefont {T.}~\bibnamefont {Ewart}}, \bibinfo {author} {\bibfnamefont
  {A.}~\bibnamefont {Kantian}}, \bibinfo {author} {\bibfnamefont
  {T.}~\bibnamefont {Giamarchi}}, \ and\ \bibinfo {author} {\bibfnamefont
  {M.}~\bibnamefont {Troyer}},\ }\href@noop {} {\bibfield  {journal} {\bibinfo
  {journal} {Computer Physics Communications}\ }\textbf {\bibinfo {volume}
  {185}},\ \bibinfo {pages} {3430} (\bibinfo {year} {2014})}\BibitemShut
  {NoStop}%
\end{thebibliography}%

\end{document}